\documentclass{sig-alternate}
\usepackage{graphicx}
\usepackage{xcolor}
\usepackage{multirow}
\usepackage{amsmath, amssymb, bbm}
\usepackage{caption}
\usepackage{subcaption}
\usepackage{balance}
\usepackage{tikz}
\usetikzlibrary{fit,positioning}
\usetikzlibrary{decorations.shapes}
\usetikzlibrary{shapes,snakes}

\tikzstyle{line}=[draw]
\definecolor{urobilin}{rgb}{0.88, 0.68, 0.13}
\definecolor{purple}{rgb}{0.5, 0.0, 0.5}
\definecolor{sinopia}{rgb}{0.8, 0.25, 0.04}
\definecolor{grey}{rgb}{0.5, 0.5, 0.5}
\definecolor{grey1}{rgb}{0.29, 0.29, 0.29}
\definecolor{grey2}{rgb}{0.71, 0.71, 0.71}
\tikzset{decorate sep/.style 2 args={decorate,decoration={shape backgrounds,shape=circle,shape size=#1,shape sep=#2}}}

\begin{document}
%

\title{Latent Space Model for Multi-Modal Social Data}
%
\def\sharedaffiliation{
\end{tabular}
\begin{tabular}{c}}

\numberofauthors{1} 
\author{
%
%
\alignauthor
 Yoon-Sik Cho, Greg Ver Steeg, Emilio Ferrara, Aram Galstyan
\sharedaffiliation
\affaddr{\small University of Southern California, Information Sciences Institute, 4676 Admiralty Way, Marina del Rey, CA 90292 USA}
}

\date{}

\maketitle
\begin{abstract}
%
With the emergence of social networking services, researchers enjoy the increasing availability of large-scale heterogenous datasets capturing online user interactions and behaviors. Traditional analysis of techno-social systems data has focused mainly on describing either the dynamics of social interactions, or the attributes and behaviors of the users. However, overwhelming empirical evidence suggests that the two dimensions affect one another, and therefore they should be jointly modeled and analyzed  in a multi-modal framework. The benefits of such an approach include the ability to build better predictive models, leveraging social network information as well as user behavioral signals.  
To this purpose, here we propose the Constrained Latent Space Model (CLSM), a generalized framework that combines Mixed Membership Stochastic Blockmodels (MMSB) and Latent Dirichlet Allocation (LDA) incorporating a constraint that forces the latent space to concurrently describe the multiple data modalities. We derive an efficient inference algorithm based on Variational Expectation Maximization that has a computational cost linear in the size of the network, thus making it feasible to analyze massive social datasets. We validate the proposed framework on two problems: prediction of social interactions from user attributes and behaviors, and behavior prediction exploiting network information.
We perform experiments with a variety of multi-modal social systems, spanning location-based social networks (Gowalla), social media services (Instagram, Orkut), e-commerce and review sites (Amazon, Ciao), and finally citation networks (Cora). 
The results indicate significant  improvement in prediction accuracy over state of the art methods, and demonstrate the flexibility of the proposed approach for addressing a variety of different learning problems commonly occurring with multi-modal social data.
\end{abstract}

\noindent \textbf{Keywords}: multi-modal social networks, topic models, LDA
\noindent \textbf{Categories and Subject Descriptors}: 
\noindent \textbf{[Networks]}: Network types---\emph{Social media networks}; \textbf{[Information systems]}: World Wide Web---\emph{Social networks}. 



\section{Introduction}
The proliferation of massive-scale social networking services produces extensive amounts of data describing various forms of user behaviors along with detailed information about social interactions. Traditionally, studies of online social systems have focused on analyzing the networks induced by such interactions \cite{adamic2005political, mislove2007measurement, huberman2008social, kwak2010twitter, mucha2010community, wu2011says, ferrara2012large, demeo2014facebook} and examining the impact of structural features on a variety of dynamical processes unfolding on the network (e.g., information diffusion, cascades, etc.) \cite{budak2011limiting, lerman2012information, cha2012delayed, goel2012structure, bakshy2012role, banos2013role, ferrara2013traveling, myers2014bursty, cheng2014can, ferrara2015quantifying}. However, a number of recent studies makes it increasingly clear that focusing only on network properties while discarding rich contextual information limits our comprehensive understanding of complex social systems dynamics  \cite{szell2010multirelational, kim2011modeling, al2012homophily, yang2013community, demeo2013analyzing, jiang2013understanding, romero2013interplay, kosinski2013private, dow2013anatomy, friggeri2014rumor, gong2014joint, agreste2015analysis}. 

In addition to social network data, we often have information about user behaviors such as mobility patterns (e.g., Foursquare check-ins), user generated content (e.g., tweets or Facebook posts), purchase history,  and so on. User behaviors and social interactions are often closely linked. This relationship has been extensively studied in social sciences under the umbrella of {\em homophily theory}~\cite{mcpherson2001birds}, which states that users exhibiting similar attributes or activity patterns are also more likely to be socially linked. When user attributes relevant to forming social links are not directly observable, this phenomenon is called {\em latent homophily}. 

From the predictive modeling perspective, homophily (or its opposite, \emph{heterophily}) can be used to build more accurate models of user behavior and social interactions based on multi-modal data. Such an approach can generate a more comprehensive understanding of users and their preferences. Furthermore, these types of models can be employed to generate predictions across modalities. For instance, we can try forecast the social interactions of an individual based on (partial) knowledge of her/his behaviors. Conversely, we can predict user behavior based on  available social network information.  This type of capability is particularly important where we have sparse data in one single modality, yet data across multiple modalities abound.  Thus, there is an overwhelming need for developing a unified computational framework to describe and analyze multi-modal social  data. 

Latent space models~\cite{Hoff01latentspace} provide a viable framework to analyze multi-modal social data. Indeed, models based on \emph{Latent Dirichlet Allocation} (LDA)~\cite{Blei:2003:LDA:944919.944937} and its extensions have been recently used to analyze user-generated content~\cite{Pang:2011:STD:1943784.1944118}, mobility patterns~\cite{Joseph:2012:BLC:2370216.2370422}, etc. At the same time, latent space models such as 
 \emph{Mixed Membership Stochastic Blockmodels} (MMSB) have been devised to describe multi-faceted interactions in social networks~\cite{Airoldi:2008:MMS:1390681.1442798}.

Using a latent space model that jointly represents information about user behaviors (or attributes) and social interactions is an attractive approach since it provides an explicit and concrete mechanism to explain the observed correlations between the two. \emph{Pairwise Link-LDA}~\cite{Nallapati:2008:JLT:1401890.1401957} combines LDA~\cite{Blei:2003:LDA:944919.944937} and MMSB~\cite{Airoldi:2008:MMS:1390681.1442798} to describe both the network structure and user-generated content. Unfortunately, Pairwise Link-LDA often fails to exploit the synergies  between different modalities, as it tends to {\em fragment} the latent space into non-overlapping (or weakly overlapping) regions corresponding exclusively to either user attributes or to the social network. This fragmentation is due to the weak dependency between the attributes and links which are connected only through a common latent topic distribution. To address this shortcoming, the \emph{Relational Topic Model} (RTM)~\cite{Chang_relationaltopic} was designed to impose additional constraints on the generative model that would require stronger correlations between links and attributes. However, RTM utilizes a significantly simplified model of link formation that does not account for truly multi-faceted interactions among the users. A detailed technical account of these existing methods and their limitations is provided in Section \S\ref{sec:related}.

In this paper, we propose an alternative approach to modeling multi-modal social data that combines the benefits of  Pairwise Link-LDA and RTM while avoiding their shortcomings. Our proposal, called \textit{Constrained Latent Space Model} (CLSM), augments the generative process by introducing a set of constraints that allows to account for stronger correlations between user attributes and social networks, and avoid the fragmentation of the joint latent space. In contrast to other methods such as RTM, CLSM uses the full generative process of MMSB to describe the link formation, which yields a truly multi-faceted modeling of social interactions. Our experiments in link and behavior prediction reveal that the latter feature is particularly important when nodes can interact with each other in multiple modes/dimensions. Overall, the proposed method achieves significant improvement in performance over the others in various learning scenarios.

Our primary contributions can be summarized as follows:
\begin{itemize}
\item We propose a general framework called Constrained Latent Space Model (CLSM) that combines MMSB and LDA with an added constraint that forces the latent space to jointly describe multiple data modalities.

\item We derive an efficient inference and learning algorithm based on \emph{Variational EM} (expectation-maximization), which makes use of the constraint to significantly reduce the computational time to a linear cost in the size of the network, allowing for large-scale data modeling.

\item We perform extensive experiments on datasets that augment social network information with various data modalities including texts, purchase reviews, and check-in data. Our framework improves the state of the art both in link prediction and in user behavior prediction.

\item We analyze two case studies in detail, the citation network \emph{Cora} and the location-based social network \emph{Gowalla}, discussing the unique insights CLSM yields over using other multi-modal latent space models.

\end{itemize}

The rest of the paper is organized as follows: Sec.~\S\ref{Prelim} introduces the background on latent space models and mentions methods related to our proposal. Sec.~\S\ref{Model} presents the Constrained Latent Space Model (CLSM), and describes the Variational Expectation Maximilazion algorithm. A rigorous experimental evaluation for link and attribute prediction is provided in Sec.~\S\ref{Exps}, with benchmarks on synthetic data and analyses of six real-world multi-modal social systems. Sec.~\S\ref{Conclude} summarizes our contribution and future plans.

\section{Background} \label{Prelim}
Consider a set of $N$ nodes that form a network  $Y$: $y(n_1,n_2) =1$ if nodes $n_1$ and $n_2$ are linked, and $y(n_1,n_2) =0$ otherwise. While the model considered here is quite general, we will focus on undirected networks where $y(n_1,n_2) = y(n_2,n_1)$. 

In addition to network information, we also have information about node behavior, such purchasing a certain item, or visiting a certain location. We assume that the behaviors are selected from a finite set, or vocabulary, ${\mathcal V }=\{1,..., V\}$. We characterize the selection $v\in {\mathcal V }$ by a vector $\mathsf{w}$ of size $V$, so that all the components of $\mathsf{w}$ are zero except the $v$-th component, which is equal to $\mathsf{w}^v=1$. Let $M_n$ be the number of times we have observed node $n$'s selections. Then the nodes's behavior is represented by $\mathbf{w}_n = ( \mathsf{w}_1, \mathsf{w}_2,..., \mathsf{w}_{M_n}  ) $.


We assume that the behavior $\mathbf{w}_n$ of user $n$ is characterized by a latent {\em topic} distribution $\boldsymbol{\theta}_n$ defined over a $K$-dimensional simplex, where $K$ is the number of topics. In the context of the present work, a topic can be interpreted as a behavior type. Each topic itself is a distribution over $V$ possible selections (locations, purchases, etc.), denoted here by $\boldsymbol{\omega}_k$, $k=1,\dots,K$. In the LDA model used here, the topics are sampled from a Dirichlet prior, $\boldsymbol{\theta}_n\sim \text{Dir}(\alpha)$. To generate the $m$-th  component of the attribute vector for user $n$, we first sample a topic indicator variable from a multinomial distribution, $\mathbf{z}_{n,m}\sim  \mbox{Mult} (\boldsymbol{\theta}_n)$, and then sample a selection $\mathbf{w}_{n,m}$ given the selected topic and its corresponding multinomial distribution: $\mathbf{w}_{n,m}\sim  \mbox{Mult} (\boldsymbol{\omega}_{z_{n,m}})$.


The main premise behind the joint latent space modeling approach is that  the same topic distribution also determines the node's social interactions, where the topics now can be interpreted as different groups or communities a node is affiliated with. Namely, for a pair of nodes $(n_1,n_2)$, we sample two indicator variables $\mathbf{z}_{n_1 \rightarrow n_2}$ and $\mathbf{z}_{n_2 \leftarrow n_1}$, and then establish a link between those two nodes with probability $f(\mathbf{z}_{n_1 \rightarrow n_2},\mathbf{z}_{n_1 \leftarrow n_2})$, where $f$ is a {\em link} function. For instance, in  Mixed Membership Stochastic Blockmodels, $f$ is parameterized via a $K\times K$ {\em compatibility} matrix $\mathsf{B}$ so that $\mathsf{B}_{k_1 k_2}$ is the probability of establishing a link between nodes from groups $k_1$ and $k_2$.   When the off-diagonal components in $\mathsf{B}$ are zero, the only allowed interactions are intra-group ones, which yields strictly \emph{assortative Mixed Membership Stochastic Blockmodels} (aMMSB)~\cite{conf/nips/GopalanMGFB12}. We distinguish  aMMSB from conventional MMSB~\cite{Airoldi:2008:MMS:1390681.1442798} by naming the latter model as \emph{full} MMSB (fMMSB), to reflect the use of a full compatibility matrix.  Throughout our experiments, we found that aMMSB better capture the joint latent space. Hence, we will use aMMSB for the Pairwise Link-LDA model as our baseline, which yields better results than that presented in previous work~\cite{Nallapati:2008:JLT:1401890.1401957} using fMMSB. 


\subsection{Related Work} 
\label{sec:related}
Before describing the proposed model, we review two related approaches that are the closest to our model, Pairwise Link-LDA~\cite{Nallapati:2008:JLT:1401890.1401957} and RTM~\cite{Chang_relationaltopic}.

Both approaches use an LDA-based process for generating behaviors. Specifically, for each user $n$, we sample topic weights $\boldsymbol\theta_n$ from a Dirichlet prior ${\boldsymbol{\alpha}}$, and a number of selections $M_n$  from a Poisson distribution. Then, for each selection $m$, we sample a topic indicator  $\mathbf{z}_{n,m}$, and generate attributes from the corresponding multinomial distribution, $\mathbf{w}_{n,m}\sim \mbox{Mult} (\boldsymbol \omega_{\mathbf{z}_{n,m}} )$.  

The two approaches differ in their link formation model. In Pairwise Link-LDA, the links are generated according to the Mixed Membership Stochastic Blockmodel~\cite{Airoldi:2008:MMS:1390681.1442798}. Namely, given two nodes $n_1$ and $n_2$, we sample indicator variables from the corresponding  topic distributions, $\mathbf{z}_{n_1 \rightarrow n_2}\sim \mbox{Mult} (\boldsymbol\theta_{n_1})$, $\mathbf{z}_{n_1 \leftarrow n_2}\sim  \mbox{Mult} (\boldsymbol\theta_{n_2})$, and then generate a link according to a Bernoulli trial, $y(n_1,n_2) \sim \mbox{Bernoulli}( \mathbf{z}_{ {n_1} \rightarrow {n_2}} ^\top  \mathsf{B} \mathbf{z}_{{n_1} \leftarrow {n_2}})$, where $\mathsf{B}$  is the $K\times K$ compatibility matrix. 



From the joint modeling perspective, Pairwise Link-LDA imposes correlations between user behavior and social interactions by sampling the corresponding indicator variables from the same (multinomial) distributions. As we mentioned above, this framework fails to properly account for observed correlations, especially when the number of topics is large~\cite{Chang_relationaltopic}. Instead, the latent space is fragmented into network-specific and behavior-specific subspaces. A similar phenomenon was observed in  \cite{Blei:2003:MAD:860435.860460} where the authors used a shared latent space for describing images and their captions.

The Relational Topic Model (RTM)~\cite{Chang_relationaltopic} tried to address this problem by modifying the link formation mechanism. Namely, instead of sampling new indicator variables for link formation, RTM {\em reuses} the indicator variables used for generating behaviors. Let  $ \bar{\mathbf{z}}_n$ be the mean of the indicator variables used for generating behaviors, $ \bar{\mathbf{z}}_n = {1\over M_n}\sum \mathbf{z}_{n,m}$. Then, given a pair of nodes $(n_1,n_2)$, a link is sampled via $y(n_1,n_2) \sim \mbox{Bernoulli}(f_{RTM}(\bar{\mathbf{z}}_{n_1} \circ \bar{\mathbf{z}}_{n_2}))$, where $\circ$ denotes the Hadamard (element-wise) product, and $f_{RTM}(\cdot)$ is a parameterized link function, e.g., sigmoidal or exponential~\cite{Chang_relationaltopic}.


While RTM is superior to Pairwise Link-LDA for describing correlations between user attributes and interactions, its simplified link formation mechanism does not have the flexibility of MMSB, and is not adequate for capturing  truly heterogenous and multi-faceted interactions between the users. Indeed, as illustrated in Section~\ref{sec:exp_synthetic}, RTM is not able to accurately model scenarios where the nodes have well-mixed memberships. We next formulate our proposed generative model that addresses the shortcomings of both methods.

\section{Constrained Latent Space Model} \label{Model} 

\begin{figure*} [t!]
        \centering
        \begin{subfigure}[t]{0.45\textwidth}
        \centering
\scalebox{0.7}{
\begin{tikzpicture}
\tikzstyle{main}=[circle, minimum size = 10mm, thick, draw =black!80, node distance = 8mm]
\tikzstyle{connect}=[-latex, thick]
\tikzstyle{box}=[rectangle, draw=black!100]
  \node[main, fill = white!100] (alpha) [label=below:$\boldsymbol \alpha$] { };
  \node[main] (pi_a) [above right= 15mm of alpha,label=below:$\boldsymbol{\theta}_n$] { };
  \node[main] (z_ab) [right= of pi_a,label=below:$\mathbf{z}_{n \rightarrow n'}$] { };
  \node[main] (pi_b) [below right= 15mm of alpha,label=above:$\boldsymbol{\theta}_{n'}$] { };
  \node[main] (z_ba) [right= of pi_b,label=above:$\mathbf{z}_{n \leftarrow n'}$] { };
  \node[main] (z_a) [above=of pi_a,label=above:$\mathbf{z}_n$] { };
  \node[main] (z_b) [below= of pi_b,label=below:$\mathbf{z}_{n'}$] { };
  \node[main, fill = black!10] (w_a) [right=25mm of z_a,label=above:$\mathbf{w}_n$] { };
  \node[main, fill = black!10] (w_b) [right=25mm of z_b,label=below:$\mathbf{w}_{n'}$] { };
  \node[main, fill = black!10] (y) [below right=15mm of z_ab,label=below:${y(n,n')}$] { };
  \node[main] (n) [left= of y,label=below:$\boldsymbol \eta$] { };

  \path (alpha) edge [connect] (pi_a)
        (alpha) edge [connect] (pi_b)
        (pi_a) edge [connect] (z_ab)
        (pi_a) edge [connect] (z_a)
        (z_a) edge [connect] (w_a)
        (pi_b) edge [connect] (z_ba)
        (pi_b) edge [connect] (z_b)
        (z_b) edge [connect] (w_b)
        (z_ab) edge [connect] (y)
        (z_ba) edge [connect] (y)
        (n) edge [connect] (y);

  \node[rectangle, inner sep=0mm, fit= (z_a) (w_a),label=below right:$M_n$, xshift=8mm] {};  
  \node[rectangle, inner sep=4.4mm,draw=black!100, fit= (z_a) (w_a)] {};
  \node[rectangle, inner sep=0mm, fit= (z_b) (w_b),label=above right:$M_{n'}$, xshift=8mm] {};  
  \node[rectangle, inner sep=4.4mm,draw=black!100, fit= (z_b) (w_b)] {};
  
\end{tikzpicture}
}
        \label{fig:a}
        \caption{}
        \end{subfigure}
        ~
        \begin{subfigure}[t]{0.45\textwidth}
        \centering
\scalebox{0.7}{
\begin{tikzpicture}
\tikzstyle{main}=[circle, minimum size = 10mm, thick, draw =black!80, node distance = 8mm]
\tikzstyle{connect}=[-latex, thick]
\tikzstyle{box}=[rectangle, draw=black!100]
  \node[main, fill = white!100] (alpha) [label=below:$\boldsymbol \alpha$] { };
  \node[main] (pi_a) [below right=of alpha,label=below:$\boldsymbol {\theta}_n$] { };
  \node[main] (z_ab) [right=of pi_a,label=above:$\mathbf{z}_{n \rightarrow \cdot}$] { };
  \node[main] (pi_b) [above right=of alpha,label=below:$\boldsymbol {\theta}_{n'}$] { };
  \node[main] (z_ba) [right=of pi_b,label=below:$\mathbf{z}_{n \leftarrow {n'}}$] { };
  \node[main] (z_a) [below=25mm  of z_ab,label=below:$\mathbf{z}_n$] { };
  \node[main] (pi_bprime) [above=of pi_b,label=below:$\boldsymbol{\theta}_{n''}$] { };
  \node[main] (z_baprime) [right=of pi_bprime,label=below:$\mathbf{z}_{n \leftarrow {n''}}$] { };
  \node[main, fill = black!10] (w_a) [right=20mm of z_a,label=below:$\mathbf{w}_n$] { };
  \node[main, fill = black!10] (y) [right=20mm of z_ab,label=below :${y(n,\cdot)}$] { };
  \node[main] (n) [below left=12mm of y,label=below:$\boldsymbol \eta$] { };

  \path (alpha) edge [connect] (pi_a)
        (alpha) edge [connect] (pi_b)
         (alpha) edge [connect] (pi_bprime)
        (pi_a) edge [connect] (z_ab)
        (z_ab) edge [connect] (z_a)
        (z_a) edge [connect] (w_a)
        (pi_b) edge [connect] (z_ba)
        (pi_bprime) edge [connect] (z_baprime)
        (z_ab) edge [connect] (y)
        (z_ba) edge [connect] (y)
        (n) edge [connect] (y)
        (z_baprime) edge [connect] (y);
        
\draw[decorate sep={0.7mm}{3mm},fill] (60pt,50pt) -- (60pt,80pt);

  \node[rectangle, inner sep=0mm, fit= (z_a) (w_a),label=above right:$M_n$, xshift=8mm] {};
  \node[rectangle, inner sep=4.4mm,draw=black!100, fit= (z_a) (w_a)] {};

  \node[rectangle, inner sep=0mm, fit= (z_ab) (y),label=above right:$N-1$, xshift=8mm] {};
  \node[rectangle, inner sep=4.4mm,draw=black!100, fit= (z_ab) (y)] {};
  
\end{tikzpicture}
}
                \label{fig:b}
                \caption{}
                
        \end{subfigure}
        \caption{\emph{Plate Diagram} comparison of Pairwise Link-LDA (a) and CLSM (b). Shaded nodes represent the observable variable. (a) Pairwise Link-LDA focusing on a pair: attributes and links are generated from the same latent variable, but a dependency between the latent factors $\mathbf{z}_{n\rightarrow n'}$ and $\mathbf{z}_n$ lacks. (b) CLSM: we focus on node $n$ and the behaviors of other nodes are omitted. Dot represents all possible nodes interacting with node $n$. $\mathbf{z}_n$ now has high dependency on the set of $\mathbf{z}_{n\rightarrow \cdot}$.   }
        \label{fig:plate} 
\end{figure*}
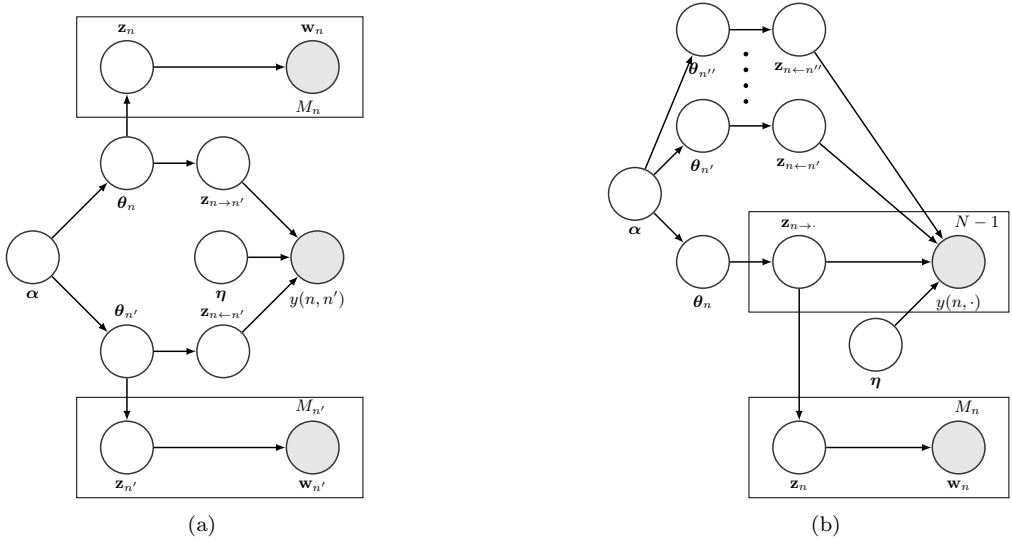

This section describes our proposal, the \emph{Constrained Latent Space Model} (CLSM).
The plate diagram of the generative process is  illustrated in Figure~\ref{fig:plate}, which for sake of comparison also depicts the generative process for Pairwise Link-LDA. Both approaches use the same scheme to generate the social network.  Where the approaches differ, however, is the generative process for the user behaviors (attributes). Namely, when generating behaviors for user $n$, we reuse the indicator variables that were used to generate links for the same user. 
In this regard, our approach shares some similarity with RTM, in that it also uses the same set of indicator variables for both attribute and link generation. In contrast to RTM, however, we first generate a set of indicator variables by sampling the links, and then use this set to generate attributes. Remarkably, our experiments show that the order of sampling is not important, and we obtain the same results by first sampling attributes, as done in RTM. However, sampling the links first is computationally more efficient when the network is relatively sparse, which is the case for the vast majority of datasets reflecting real-world techno-social systems. In the opposite case of dense networks and sparse attributes,  one can sample the other way around  for efficiency. 

Following the language of MMSB,  let $N$ be the total number of nodes, $K$ be the total number of network communities (that is, topics). The generative process is as follows: 
\begin{itemize}
\item {\bf Network Generation:} 
\begin{enumerate}
\item For each community $k$, sample the community strength $\beta_k \sim \mbox{Beta} (\eta_1, \eta_0)$, where $\beta_k := \mathsf{B}(k,k)$
\item For each node $n$, sample the $K \times1$ membership vector $\boldsymbol{\theta}_n \sim \mbox{Dirichlet}( {\boldsymbol{\alpha}})$. 
\item For each node $n$, initialize an empty set of indicator variables, $\mathbb{Z}_n=\emptyset$. 

\item For each pair ($n_1$, $n_2$)
\begin{enumerate} 
\item Draw membership indicator vectors \\
$\mathbf{z}_{{n_1} \rightarrow {n_2}}  \sim \mbox{Mult}( \boldsymbol{\theta}_{n_1})$\\
$\mathbf{z}_{{n_1} \leftarrow {n_2}}  \sim \mbox{Mult}( \boldsymbol{\theta}_{n_2})$.
\item Sample the pair interaction \\ 
$y(n_1,n_2) \sim \mbox{Bernoulli}( \mathbf{z}_{ {n_1} \rightarrow {n_2}} ^\top  \mathsf{B} \mathbf{z}_{{n_1} \leftarrow {n_2}})$, \\
where $\mathsf{B} = \mbox{diag} (\beta_1,\beta_2,...,\beta_K) +  \epsilon(\mathbbm{1}-I)$.
\item Augment the corresponding indicator sets \\
$\mathbb{Z}_{n_1}\rightarrow \mathbb{Z}_{n_1} \cup \{\mathbf{z}_{{n_1} \rightarrow {n_2}}\}$ \\
$\mathbb{Z}_{n_2}\rightarrow \mathbb{Z}_{n_2} \cup \{\mathbf{z}_{{n_1} \leftarrow {n_2}}\}$

\end{enumerate}
\end{enumerate}
\item {\bf Behavior Generation:} 

Let $M_n$ be the total number of selections\footnote{For the purposes of data generation $M_n$ can be sampled from, say, a Poisson distribution. This is not relevant for inference, however, where $M_n$ is specified in the data.} of user $n$ from a (behavior) set ${\mathcal V }=\{1,..., V\}$. 
\begin{enumerate}
\item For each topic $k$, sample the attribute distribution vector $ \boldsymbol{\omega}_k  \sim \mbox{Dirichlet} (\boldsymbol{\kappa})$.
\item For each selection $\mathbf{w}_{n,m}$ of user $n$, where $m \in \{ 1,..., M_n\}$, 
\begin{enumerate}
\item Sample an integer $\mathbf{c}_m^n \sim \mbox{Unif}( \{1,..., \text{size}(\mathbb{Z}_n)  \}  )$, and let  $\hat{\mathbf{z}}$ be the topic indicator vector corresponding to $\mathbf{c}_m^n$.
\item Sample a selection $\mathbf{w}_{n,m} \sim \mbox{Mult} (\boldsymbol\omega_{\hat{\mathbf{z}}} )$,  \\
\end{enumerate} 
\end{enumerate} 
\end{itemize}
Exact inference with the above generative model is not feasible due to the presence of latent variables~\cite{Hoff01latentspace, blei2012probabilistic}. In the following, we describe an approximate method for learning and inference  based on Variational Expectation Maximization (EM), which has a computational cost linear in the size of the network. The adoption of this approach allows using CLSM with real-world (possibly very large) multi-modal social datasets.

\subsection{Variational Inference} 
Given the observation of the interaction network $\mathsf{Y}$ and user behaviors $\mathbf{w}_{1:N}=\{\mathbf{w_k}\}_{k=1}^N$, we are interested in inferring the posterior distributions of the model's latent variables, $p( \boldsymbol{\theta}_{1:N}, \mathbb{Z}_{1:N}, \mathbf{C}|\mathsf{Y}, \mathbf{w}_{1:N})$ (where $\mathbf{C}$ is the collection of all $\mathbf{c}_m$ of all nodes),  as well as  estimating the hyperparameters $\boldsymbol{\eta}, \boldsymbol{\alpha}, \boldsymbol{\kappa}$.  A number of approximate inference algorithms have been proposed in literature. In this paper, we rely on variational method~\cite{VI} that approximates the posterior by a computationally tractable variational distribution with some free variational parameters. Those parameters are selected to minimize  the Kullback-Leibler divergence between the variational distribution and the true posterior. 

Here we suggest a factorized  variational   distribution over the latent variables $q( \boldsymbol{\theta}_{1:N}, \mathbb{Z}_{1:N}, \mathbf{C})$: 
\begin{eqnarray}
q( \boldsymbol{\theta}_{1:N}, \mathbb{Z}_{1:N}, \mathbf{C}) = \prod_{n=1}^N q_{\text{dir}}( \boldsymbol{\theta}_n | \boldsymbol{\gamma}_n) 
\prod_{m=1}^{M_n} q_{\text{mul} } (\mathbf{c}_m^n | \boldsymbol{\lambda}_m^n) \nonumber \\
 \prod_{n_1,n_2}^N q_{\text{mul} }(\mathbf{z}_{n_1 \rightarrow n_2}  |\boldsymbol{ \phi} _{n_1 \rightarrow n_2}) q_{\text{mul} }(\mathbf{z}_{n_1 \leftarrow n_2}  | \boldsymbol{\phi} _{n_1 \leftarrow n_2}). 
\end{eqnarray}   
 where $\{\boldsymbol{\gamma}\}$, $\{\boldsymbol\lambda\}$, and $\{\boldsymbol\phi\}$ are the variational parameters. Similarly, for the distributions over the model's parameters, we also use factorized $q(\cdot)$ distributions, which are $q_{\text{beta}}(\beta_k | \tau_{k1}, \tau_{k0}  )$ and 
$q_{\text{dir}}( \boldsymbol{ \omega}_{i} | \boldsymbol{\rho}_{i} )$. 

Given the factorized variational distribution $q(\cdot)$, we next bound the log likelihood of the observed data  using Jensen's inequality. Specifically, we consider the so called evidence lower bound (ELBO)  defined as follows: 
\begin{eqnarray}
\log p(\mathsf{Y}, \mathbf{w}_{1:N} |\boldsymbol{\eta}, \boldsymbol{\alpha}, \boldsymbol{\kappa} ) \geq \mathcal{L}(\boldsymbol{\phi}, \boldsymbol{\gamma},\boldsymbol{\lambda}) ~~~~~~~~~~~~~~~~~~~~~~~~~~~~~~\nonumber \\  
 \triangleq \mathbb{E}_q[\log p( \mathsf{Y}, \mathbf{w}_{1:N}, \boldsymbol{\theta}_{1:N}, \mathbb{Z}_{1:N}, \mathsf{B}, \mathbf{C}, \mathsf{\Omega} |\boldsymbol{\eta}, \boldsymbol{\alpha}, \boldsymbol{\kappa}   )] \nonumber \\ 
- \mathbb{E}_q[\log q( \boldsymbol{\theta}_{1:N}, \mathbb{Z}_{1:N}, \mathsf{B}, \mathbf{C}, \mathsf{\Omega}) ]  ,   \label{eq:ELBO}
\end{eqnarray}
where we defined $\mathsf{\Omega}$ as a collection of $\boldsymbol{\omega}$.

The ELBO in Equation~\ref{eq:ELBO} is expanded as follows: 
\begin{align} 
\mathcal{L} ={}& \sum_{n_1,n_2} \mathbb{E}_q [   \log p(y(n_1,n_2) | \mathbf{z}_{n_1 \rightarrow n_2} , \mathbf{z}_{n_1 \leftarrow n_2} ,  \boldsymbol{\beta} ]   \nonumber \\
	+{}& \sum_{n_1,n_2} \mathbb{E}_q [ \log p( \mathbf{z}_{n_1 \rightarrow n_2} | \boldsymbol{\theta}_{n_1}  ) + \log p( \mathbf{z}_{n_1 \leftarrow n_2} | \boldsymbol{\theta}_{n_2}  ) ] \nonumber\\
         +{}& \sum_{n} \sum_{m=1:M_n}  \mathbb{E}_q [ \log p(\mathbf{w}_{n,m} )  | \mathbf{c}_m^n),\mathsf{\Omega} ]      \nonumber\\
         +{}& \sum_{n} \sum_{m=1:M_n}  \mathbb{E}_q [ \log p(\mathbf{c}_m^n  | \mathbf{z}_{n \rightarrow \cdot}, \mathbf{z}_{\cdot \leftarrow n} ) ]      \nonumber\\
	+{}& \sum_n  \mathbb{E}_q [\log p(\boldsymbol{\theta}_n  | \boldsymbol{\alpha} ) ]   \nonumber\\  
	+{}& \sum_k  \mathbb{E}_q[    \log p(\boldsymbol{\beta}_k | \boldsymbol{\eta}) ] +  \sum_k  \mathbb{E}_q[  \log p( \boldsymbol{\omega}_k | \boldsymbol{\kappa}) ] \nonumber\\
	-{}& \sum_{n_1,n_2}  \mathbb{E}_q [ \log q( \mathbf{z}_{n_1 \rightarrow n_2} | \boldsymbol{\phi}_{n_1 \rightarrow n_2}  ) + \log q( \mathbf{z}_{n_1 \leftarrow n_2} | \boldsymbol{\phi}_{n_1 \leftarrow n_2}  ) ] \nonumber\\
	-{}& \sum_n\sum_{m=1:M_n} \mathbb{E}_q[ \log q(\mathbf{c}_m^n | \boldsymbol{\lambda}_m^n ) ]  \nonumber\\
	-{}& \sum_n \mathbb{E}_q[ \log q(\boldsymbol{\theta}_n | \boldsymbol{\gamma}_n)] .
\end{align} 

The lower bound can be maximized using the coordinate ascent algorithm. Toward that goal, we take the (partial) derivatives of $\mathcal{L}(\boldsymbol{\phi}, \boldsymbol{\gamma},\boldsymbol{\lambda})$ with respect to the variational parameters and set them to zero. As a result, we obtain a set of iterative update equations for the parameters, whose fixed points correspond to local optima of the objective function. 

For $\boldsymbol{\gamma}_n$, the update equations are as follows:
\begin{equation}
\boldsymbol\gamma_{n,k} \leftarrow \boldsymbol\alpha_{k} + \sum_{n' \neq n}  \boldsymbol\phi_{n \rightarrow n' ,k} + \sum_{n' \neq n}  \boldsymbol\phi_{n' \leftarrow n ,k}. \label{eq:gamma} 
\end{equation} 
Similar reasoning leads to update equations for the parameters $\{ \boldsymbol{\phi} \}$ and $\{ \boldsymbol{\lambda} \}$, with  additional constraints that each component of those vectors sum to 1. As we mentioned above, here we are interested in undirected social interactions, which reduces the number of parameters due to the equalities $\boldsymbol{\phi}_{n \leftarrow n'} =  \boldsymbol{\phi}_{n' \rightarrow n}$. Thus, we consider only the update equations for $\boldsymbol{\phi}_{n \rightarrow n'}$. 

After adding the appropriate Lagrange multipliers to ensure the normalization of the  components, $\sum_{n'} \boldsymbol{\phi}_{n \rightarrow n'}=1$, we isolate the terms that contain $\boldsymbol{\phi}_{n \rightarrow n'}$, and set their derivatives with respect to $\boldsymbol{\phi}_{n \rightarrow n'}$ to zero. The corresponding update equations are as follows: 
\begin{eqnarray}
\boldsymbol{\phi}_{n \rightarrow n', k} \propto \exp( \mathbb{E}_q[\log p( \boldsymbol\theta_{n,k})]  ~~~~~~~~~~~~~~~~~~~~~~~~~~~~~~~~~ \label{eq:update1}\nonumber \\
+ \mathbb{E}_q[\log p(y(n,n')) ] + \mathbb{E}_q[\log \prod_{m=1}^{M_n} p(\mathbf{w}_{n,m})^{\mathbf{c}_{m,n'}^n} ]),   \label{update:phi}  
\end{eqnarray}
\begin{equation}
\boldsymbol\lambda_{m,n'}^n \propto \exp(\mathbb{E}_q[\log p( \mathbf{w}_{n,m} | \mathbf{c}_{m,n'}^n =1 )]),  
\end{equation}
where the term $n'$ appearing in $\mathbf{c}_{m,n'}^n$ and $\boldsymbol\lambda_{m,n'}^n$ denotes the index of $\boldsymbol{\phi}_{n \rightarrow n'}$ in  $\mathbb{Z}_n$. As with the indicator vector $\mathbf{z}_{n \rightarrow n'}$,  $\mathbf{c}_m^n$ should have only one component equal to 1 setting all others to 0. 

Note that the number of the parameters $\boldsymbol{\phi}_{n \rightarrow n'}$, and thus the computational complexity of the update equations for Eq.~\ref{eq:update1}, is quadratic in the number of nodes, even when the network is sparse.  This is because the parameters are defined both for links and non-links.   To avoid this computational bottleneck, we adopt an approximation technique~\cite{conf/nips/GopalanMGFB12}, by assuming that the parameters $ \boldsymbol{\phi}_{n \rightarrow \cdot}$ for non-links can be replaced by a single mean-field parameter. Namely, let  $\{ \boldsymbol{\phi}_{n \rightarrow \cdot } \}^+$ and $\{ \boldsymbol{\phi}_{n \rightarrow \cdot } \}^-$ be the set of parameters for links and non-links, respectively, and let ${\bar{\boldsymbol{\phi}}}_{n \rightarrow \cdot} $ be the average over the set $\{ \boldsymbol{\phi}_{n \rightarrow \cdot } \}^+$. Within the above mean field approximation, each element of $\{ \boldsymbol{\phi}_{n \rightarrow \cdot } \}^-$ is replaced by ${\bar{\boldsymbol{\phi}}}_{n \rightarrow \cdot} $. 


Further gains in computational efficiency is achieved by limiting the number of components in set $\mathbb{Z}_n $ to the number of edges incident on node $n$, rather than having all the relations in the set. This corresponds to reusing indicator variables only when they have generated a link.  
Note also that now Equation~\ref{eq:gamma} does not require all the parameters $\{ \boldsymbol{\phi} \}$, as the parameters correspodning to non-links are replaced by  $\bar{\boldsymbol{\phi}}_{n \rightarrow \cdot} $.

Finally, the update Equation~\ref{update:phi} can be further simplified by making the best use of the assortative property in aMMSB, where only diagonal components of the block matrix are being considered. Instead of updating $K \times K$ combinations of  $\boldsymbol{\phi}_{n_1 \rightarrow n_2}$ and  $\boldsymbol{\phi}_{n_2 \rightarrow n_1}$    for a link between node $n_1$ and $n_2$, we can update only $K$ parameters of  by disregarding the inter-community links. For our aMMSB model, we use the following update equation:  
\begin{eqnarray}
~~~~ \boldsymbol {{\phi}}_{n_1\rightarrow n_2, k} \propto \exp( \mathbb{E}_q[\log p(\boldsymbol{\theta}_{n_1,k})]+ \mathbb{E}_q[\log p(\boldsymbol{\theta}_{n_2,k})]  \nonumber \\ \label{update:phi_undirected}  
+ \mathbb{E}_q[\beta_k]  + \mathbb{E}_q[\log \prod_{m=1}^{M_{n_1}} p(\mathbf{w}_{n_1,m})^{\mathbf{c}_{m,n_2}^{n_1}} ]     ),  
\end{eqnarray}
where we further use $\mathbb{E}_q[\log p(\boldsymbol{\theta}_{n,k})] = \psi(\boldsymbol\gamma_{n,k}) - \psi(\sum_r \boldsymbol\gamma_{n,r} )$, $ \mathbb{E}_q[\boldsymbol\beta_k]  = \psi(\boldsymbol\eta_{k,1} ) -\psi(\boldsymbol\eta_{k,2}) $ employing the exponential family distribution property. As for the last term in Equation~\ref{update:phi_undirected},  we use the following equation: 
\begin{eqnarray}
 \mathbb{E}_q[\log \prod_{m=1}^{M_{n_1}} p(\mathbf{w}_{n_1,m})^{\mathbf{c}_{m,n_2}^{n_1}} ] ~~~~~~~~~~~~~~~~~~~~~~~~~~~~~~~~~~~~~~  \nonumber \\
= \sum_{m=1}^{M_{n_1}} \lambda_{m,n_2}(\psi(  \sum_r \bold{1} (\mathbf{w}_{n_1,m} = i)   \rho_{k,r  }) - \psi(\sum_i \rho_{k,i}) ). \nonumber\\  \label{eq:null}
{}
\end{eqnarray}


Let us now focus on variational distributions over the model parameters $\boldsymbol{\beta}$ and $\boldsymbol{\omega}_k$, $k=1,..,K$. We had previously defined a Beta distribution $q_{\text{beta}}(\beta_k | \tau_{k1}, \tau_{k0}  )$ with variational parameter $\boldsymbol{\tau}_1, \boldsymbol{\tau}_0$, and a Dirichlet distribution $q_{\text{dir}}( \boldsymbol{ \omega}_{i} | \boldsymbol{\rho}_{i} )$ with variational parameter $\boldsymbol{\rho}_k$. Here we omit the derivation details and only present the final update equations for these parameters:
\begin{eqnarray} 
\boldsymbol\tau_{k,1} &\leftarrow& \boldsymbol\eta_1 + \sum_{ (n_1,n_2) \in \text{link} } \boldsymbol{\phi}_{n_1\rightarrow n_2,k},   \\
\boldsymbol\tau_{k,2} &\leftarrow& \boldsymbol\eta_2 + \sum_{ (n_1,n_2) \in \text{non-link} }\boldsymbol\phi_{n_1 \rightarrow n_2,k}  \boldsymbol\phi_{n_2 \rightarrow n_1,k}, \nonumber    \\
\boldsymbol\rho_{ij} &\leftarrow&  \boldsymbol\kappa_j + \sum_{n=1}^N \sum_{m=1}^{M_n}    \bold{1} (\mathbf{w}_{n,m} = j)   \sum_{n'\in{\text{link}}} \boldsymbol\phi_{nn',i} \boldsymbol\lambda_{m,n'}.   \nonumber \\ \label{eq:rho}
{}  
\end{eqnarray}  


One the variational parameters are found, we can use them to estimate the model parameters themselves. We note that as an alternative approach,  one can also derive explicit update equations for the parameter $\boldsymbol{\omega}_i$ directly, without using the variational parameter $\boldsymbol{\rho}_i$ in Equation~\ref{eq:rho}.  $\boldsymbol{\omega}_i$ can be optimized by introducing a Lagrange multiplier where we have the update equation as follows: 
\begin{equation}
\boldsymbol{\omega}_{ij} \propto  \sum_{n=1}^N \sum_{m=1}^{M_n}    \bold{1} (\mathbf{w}_{n,m} = j)   \sum_{n'\in{\text{link}}} \boldsymbol\phi_{nn',i} \boldsymbol\lambda_{m,n'}. 
\end{equation} 
However, for this case, extra caution is needed that guarantees non-zero entities in any $\boldsymbol{\omega}_i$. This can be easily achieved by using smoothing techniques such as Laplace smoothing or pseudo-count smoothing~\cite{Jurafsky:2000:SLP:555733}. Our experimental results reflect the latter smoothing technique.

\section{Experiments}   \label{Exps} 
In this section we perform an extensive experimental evaluation of the Constrained Latent Space Model (CLSM). 
We first discuss some experiments performed with synthetic data, designed to better understand the main differences between our method and the Relational Topic Model (RTM). Then, we carry out an extensive experimental analysis using a number of real-world datasets from various techno-social systems. After describing the characteristics of these datasets, we present two problems, link and behavior prediction, and demonstrate the superiority of the proposed model over existing methods. 
We conclude discussing two case studies and the unique insights yielded by CLSM's use.

\subsection{Experiments with Synthetic Data}
\label{sec:exp_synthetic}
The main advantage of CLSM over RTM is that our model uses the aMMSB whereas RTM employs a simplified mechanism for the link generation process. We thus expect that CLSM will prove effective in presence of nodes with well-mixed memberships (or topics). By incorporating aMMSB, our model can better infer mixed topics while relying less on attribute information, if compared to RTM which heavily relies on users' attributes for accurate inference. In fact, in MMSB each  interaction is captured to describe the topic distribution of a given node, while RTM simply posits a single distribution and tries to describe every interaction of a given node with that single distribution.

Figure~\ref{fig:syn} summarizes the experiments that confirm this intuition. Let us consider nodes with two topics where some of the nodes are exclusively assigned to one topic (or community) and others are well mixed between the two.\footnote{We set Dirichlet prior $\boldsymbol{\alpha} = [ .3 ~~.3]^\top $} In real social systems, two distinct  communities can exhibit similar or very different behaviors. To reflect this, we perform two experiments: in the former the two communities are well separated (Figure~\ref{fig:syn_gen}, left panel), while in the latter we impose some arbitrarily-tuned degree of similarity between them (Figure~\ref{fig:syn_gen}, right panel).  The community-to-behavior distribution (equivalent to  the per-topic word distribution) for the two cases has been controlled as in Figure~\ref{fig:syn_gen}. 

To measure the performance of CLSM and RTM, in Figure~\ref{fig:syn_result} we plot the \emph{mean absolute error} (MAE) \cite{hyndman2006another} of reconstruction against the gap between the two distribution peaks (the larger, the least similar are the two distributions).
In the experiments, we generated synthetic data for 2,500 nodes and 1,500 behaviors (words) with two communities (topics) using the generative process of RTM. 
Figure~\ref{fig:syn_result} shows that, as the gap between the two per-topic word distributions tend to close, the performance of RTM starts to degrade, whereas the performance of our model remains substantially unchanged. 

Further inspection revealed that, when the two topic distributions become similar one other, RTM tends to assign every node to the two extreme topics (i.e., pure \texttt{topic1} or pure \texttt{topic2}), even when the nodes  actually exhibit a clear mixture of the two topics.  This causes the RTM's higher MAE when the two word distributions get similar, whereas CLSM consistently infers topics in line with the ground truth for all nodes including those exhibiting well-mixed topics. 

\begin{figure} [t!] 
        \centering
        \begin{subfigure}[b]{0.48\textwidth}
                \includegraphics[width=\textwidth,clip=true,trim=35 0 35 0]{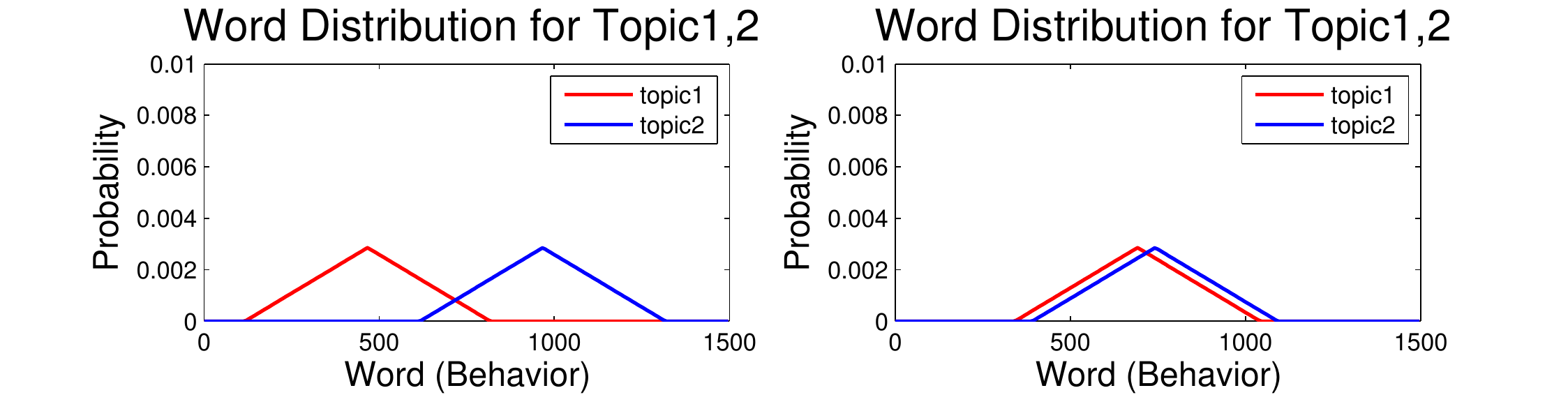}
                \caption{}
                \label{fig:syn_gen}
        \end{subfigure}
        \begin{subfigure}[b]{0.3\textwidth}
                \includegraphics[width=\textwidth,clip=true,trim=10 0 50 0]{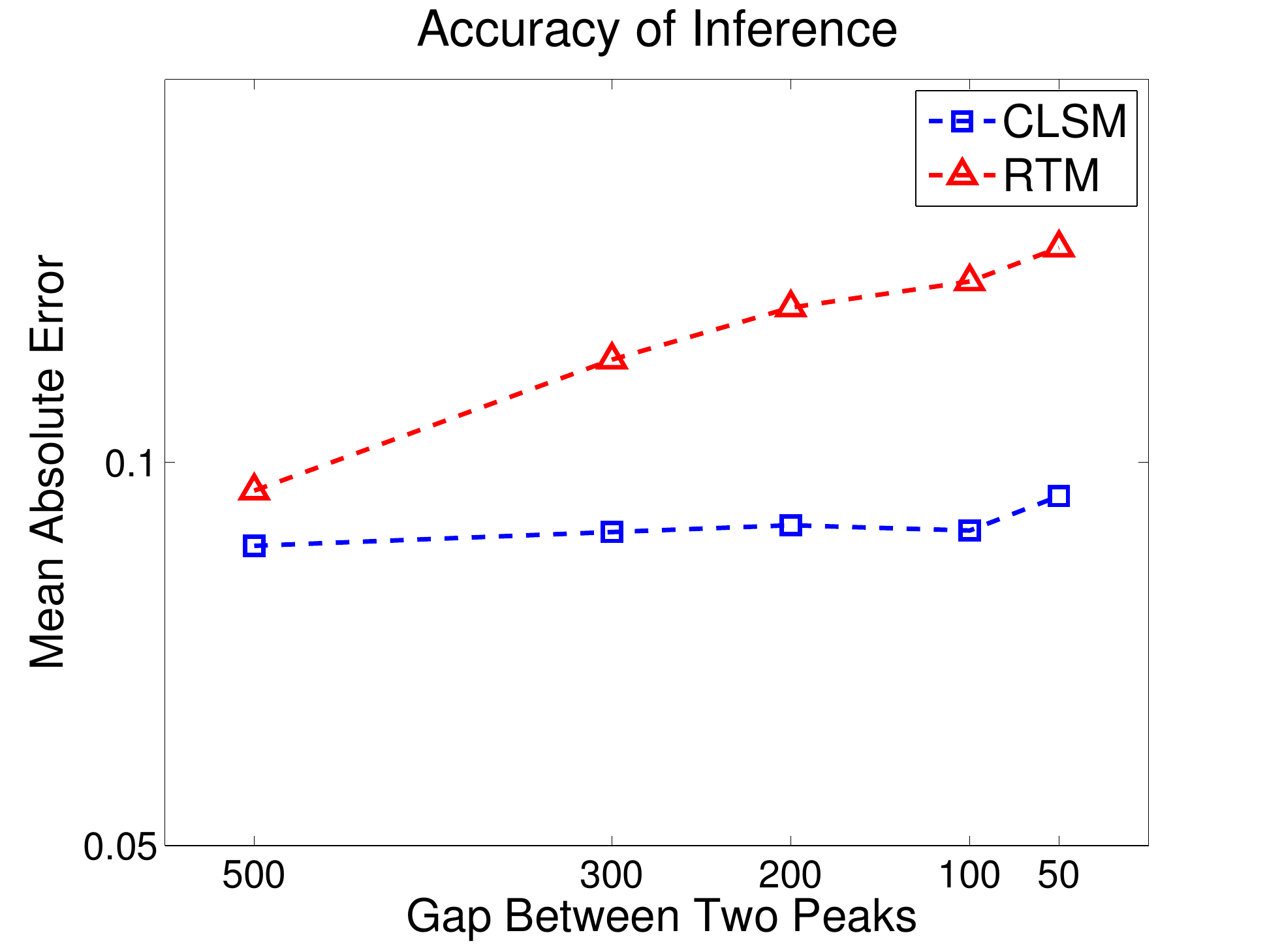}
                \caption{}
                \label{fig:syn_result}
        \end{subfigure}
        \caption{(a) Representation of two topics as distribution over the words: the left and right panels indicate weakly-overlapping and strongly-overlapping topics, respectively. The difference between the topics can be measured by the distance between their respective peaks.  (b) Mean absolute error of inference as a function of topic overlap (measured by the distance between the peaks).}
        \label{fig:syn} 
\end{figure}

\subsection{Experiments with Real-World Data} 
In the following we present experiments with various real-world  datasets. 
Our aim is twofold: first and foremost, we want to illustrate the flexibility of our framework to adapt and perform well within various learning tasks commonly occurring when analyzing multi-modal techno-social systems; second, we want to prove that CLSM outperforms the state of the art. 
We will use data from diverse types of social systems, spanning various social networks, review sites, purchasing platforms, etc. The datasets are presented in~\S\ref{dataset}.

\subsubsection{Setting the Performance Evaluation}
In our experiments, we evaluate our model against RTM and Pairwise Link-LDA on two tasks: link prediction (\S\ref{sub:lp}), and attributes prediction (\S\ref{sub:ap}). We set Pairwise Link-LDA to use aMMSB (rather than fMMSB as in the original paper~\cite{Nallapati:2008:JLT:1401890.1401957}) as this yields better accuracy, especially in the link prediction task. 
Through this fair comparison, we show that the improved performance of our method is not only the result of using aMMSB for mixed memberships, which consistently outperforms the original fMMSB.  

For both prediction tasks, we follow the settings in \cite{Chang_relationaltopic} and employ a 5-fold cross validation approach: we use the 4/5 of the data to train our models and then test their performance using the other 1/5. 
For link prediction, we hide all the links attached to the nodes in the test set, and compute the probability of a link from all the nodes in the test set to all the nodes in the training set. Similarly, for attribute prediction, we hide all the attributes of nodes in the test set, and compute the probability of each attribute. For all of the experiments, we keep the precision of the Dirichlet prior $\boldsymbol{\alpha}$ to $1$, and run the Variational EM algorithm until its convergence. For the convergence criterion, the algorithm stops when the proportional change in the likelihood bound is less than $1\text{e}-8$.

\begin{table*}[t]
\small
\caption{Summary statistics of the six multi-modal social datasets adopted for analysis and performance evaluation.}
\label{sample-table}
\begin{center}
\begin{tabular}{lcccc }
\multicolumn{1}{c}{\bf Dataset}  &\multicolumn{1}{c}{\bf {\# of Nodes}}  &\multicolumn{1}{c}{\bf { Size of Attributes}}  &\multicolumn{1}{c}{\bf {\# of Behaviors }} &\multicolumn{1}{c}{\bf {\# of Links}} 
\\ \hline \\
{\it Gowalla} (San Francisco)         &931 \small{\texttt{users}} &1,909 \small{\texttt{venues}}& 63,543 \small{\texttt{ check-ins}} & 3,429 \small{\texttt{friendships}}\\
{\it Cora}                  &2,708 \small{\texttt{documents}} & 1,433 \small\texttt{terms} & 49,216 \small\texttt{word counts} & 5,278 \small\texttt{citations}\\
{\it Ciao}                 &  1,442 \small{\texttt{users}}    &  2,238 \small\texttt{products}   &  46,732 \small\texttt{reviews}  &  29,040 \small\texttt{user-trusts} \\
{\it Amazon}                  &  2,942 \small\texttt{products}   &  1,810 \small\texttt{labels}   &  29,669 \small\texttt{descriptions}  &  18,814 \small\texttt{co-purchases}\\
{\it Orkut}                  &  3,201 \small\texttt{users}   &  2,573 \small\texttt{groups}   &  26,658 \small\texttt{affiliations}  &  47,871 \small\texttt{friendships}\\
{\it{Instagram}}	&1,818	\small\texttt{users} &	500 \small\texttt{media} &	91,246 \small\texttt{tags} &  36,342 \small\texttt{friendships} \\
\end{tabular}
\end{center}
\end{table*}

\subsubsection{Description of the Datasets} \label{dataset}
We here present the various datasets employed for performance evaluation. Some descriptive statistics are summarized in Table~\ref{sample-table}. All datasets provide a social network dimension. As for the attributes, {\it Gowalla} contains value counts, since the system allows for repeated behaviors (\emph{check-in}s) on the same attribute (venue). All the other datasets have binary representations over the attributes, which capture their presence (0 or 1). 
It's worth noting that, although the sheer size of these datasets is not massive due to the challenge of finding multi-modal social datasets for which both social network and user behaviors are available for large sets of users, we stress again how the low computational cost of CLSM makes it very suitable for large scale analysis.

\smallskip\textbf{Location-Based Social Network.} \emph{Gowalla} is a location-based social network  which allows users to {\it check-in} their current location using their mobile devices and share that with their friends. Cho \emph{et al.} \cite{FM} collected the  Gowalla check-in and social network data from Feb. 2009 to Oct. 2010. Each check-in datapoint consists of user ID, venue ID, timestamp, and location (latitude/longitude coordinates). In our experiments, we focused on the most represented US city: San Francisco. We consider users as nodes, friendships as links, and the number of  check-ins on a given venue as attributes. Note that we discard any temporal information about the check-ins, and we focus on the top active $20\%$ users, which yield over $80\%$ of the total  check-in records. 

\smallskip\textbf{Citation Network.} The {\it Cora} dataset~\cite{mccallum00automating,sen:aimag08} was the largest citation network  used to benchmark RTM's performance \cite{Chang_relationaltopic}. It contains the abstracts from the Cora research paper search engine, where the documents in this repository cite each other. We consider documents as nodes, citations as links, and the set of lexical terms as set of  attributes. We adopt the pre-processed dataset used by Chang and Blei~\cite{Chang_relationaltopic} where the observed attributes for each node convey the presence ($0$ or $1$) of a given term in the paper abstract.

\smallskip\textbf{Review-Trust Network.} This dataset consists of reviews on various products spanning DVDs to cars, and the trust network among \emph{Ciao}'s users. Each review consists of a user (or reviewer), the product, the category of the product being reviewed, the review score, and the score that measures the helpfulness. To simplify, we disregard review and helpfulness scores, and assume that the presence of a review (regardless of the score) indicates a certain amount of interest in the product by the reviewer (e.g., which led the reviewer to  buy the product and review it). Each user establishes a directed trust link with other users whom they want to follow. We treat those trust relationships as links, and the  list of reviewed products as attributes. 

\smallskip\textbf{Co-purchase Network.} In this dataset, an undirected edge describes the ``customers who bought this item also bought'' relationship. We consider each product as node in our model. Each product also has a label description as a category,  provided by {\it Amazon}. Most items are matched to multiple  hierarchical label descriptions that correspond to the product.  For instances, many digital cameras have multiple labels with \texttt{Electronics> Camera \& Photo> Digital Cameras}, and are often purchased together with memory cards, tripods, or batteries. Here, we use these label descriptions as attributes. 

\smallskip\textbf{Social Network with Special Interest Groups.} \textit{Orkut} is a social network service owned by Google. The dataset has been originally collected by Mislove \emph{et~al.}~\cite{mislove2007measurement} during a crawl performed during Oct-Nov 2006. Each user in \textit{Orkut} makes online friends with others and can also join \emph{special interest groups}. 
Users are allowed to join as many groups as they would like. A group might consist of colleagues, celebrity fans, etc. Users can make new friends in a group, thus two users being in a same group does not necessarily imply that they are friends each other. 
Here, we  consider the top 5,000 communities with the highest quality as identified by Yang and Leskovec~\cite{YangL12}. Users with no connections to these communities have been filtered out. The dataset has been further refined by excluding the users with no links, and removing the groups with no nodes of our interest. In our experiments, we use the group affiliations as attributes and the friendships between users as links.

\smallskip\textbf{Media-Sharing Social Network.} \emph{Instagram} is a social multi-media sharing platform owned by Facebook. This dataset has been collected during Jan-Feb 2014, starting from 2,100 randomly-selected user seeds who participated to at least one of 72 popular photography contests, as identified by particular hashtags \cite{ferrara2014online}. The social network mode reports follower relationships, whereas the behaviors represent the tags adopted by each user to label  photos (attributes) chosen to participate to a given contest. 

\subsubsection{Task 1: Link Prediction} \label{sub:lp}

Here we discuss our first evaluation task, link prediction. Accurate link prediction can be very useful in many real-world applications: for instance, in a location-based social network, we can use such tool to recommend friends to new users who have just joined the service and have not yet made any friends but already left some records of their activity. 

Borrowing the language of topic modeling, using the four folds of data for training, \emph{(i)} we first infer the topic distribution using the attributes, \emph{(ii)} then compute the probability of the links between the nodes,  and \emph{(iii)} finally rank the pairs according to their likelihoods. These sorted candidates are compared with the ground truth, represented by the fifth fold left out for testing purposes. Our performance evaluation measure is the \emph{average ranking score}. Figure~\ref{fig:link} shows the average ranking of positive edges with $100\%$ recalls.  Lower average ranking scores imply better performance: a user would find their ``real'' friends higher up in the recommendation list. For each dataset, we increase the number of topics from 5 to 25 by increment of 5, and for each configuration we perform 10 rounds of cross validation and average the results. We also compare the performance of the benchmarked models using the \emph{Area Under the ROC Curve} (AUC): the results are reported in Table~\ref{tb:AUC}. 

For illustration purpose, let us discuss \emph{Gowalla} first. Our approach performs better than the baselines (Pairwise Link-LDA and RTM) for any number of topics. As the number of topics increases, Pairwise Link-LDA fails to capture the joint topic space, whereas CLSM shows improvements in performance. Interestingly, when the number of topic is small, Pairwise Link-LDA performs better than RTM. We believe that a small number of topics functions as a constraint that enforces the two topic spaces to be close each other. CLSM clearly outperforms the baselines in terms of AUC scores, yielding a prediction accuracy of about 70\%, a relative improvement of 15-18\% over RTM and Pairwise Link-LDA.

We repeated the same experiments on the other datasets as well. Our model performs better on the link prediction task on every dataset and for all choices of topic sizes (see Figure~\ref{fig:link}).  The link prediction results for Pairwise Link-LDA in {\it Cora} are better than those reported in previous studies \cite{Nallapati:2008:JLT:1401890.1401957} as we take advantage of aMMSB.  Overall, our model along with RTM achieves better results as the number of topic increases --- except for the {\it Ciao} dataset which exhibits an atypical pattern (the prediction accuracy gets worse as the number of topic increases). We believe this is due to the unique characteristic of the network in this platform, namely that a link reflects whether or not the user trusts another user as a reviewer. Many factors besides the topic we inferred may play a role in trust, such as the reputation of a reviewer: it is natural to trust a reviewer who submits compelling reviews although not necessarily sharing one's same interests. Otherwise, CLSM is the only model to consistently improve in performance as the number of topics increases.

In terms of AUC scores, CLSM yields performance around 75\% for \emph{Instagram}, \emph{Cora} and \emph{Ciao}, and better than 90\% with \emph{Amazon} and \emph{Orkut}, exhibiting consistent and significant improvements. \emph{Orkut} data benefit the least from CLSM with a modest  2.2\% improvement, whereas \emph{Instagram} data boast a remarkable 22\% relative increment. For link prediction, CLSM yields an average increase across the six datasets of 12.56\% and 9.06\% respectively over Pairwise Link-LDA and RTM.

\begin{figure}[!t]
\begin{center}
 \includegraphics[width=0.234\textwidth]{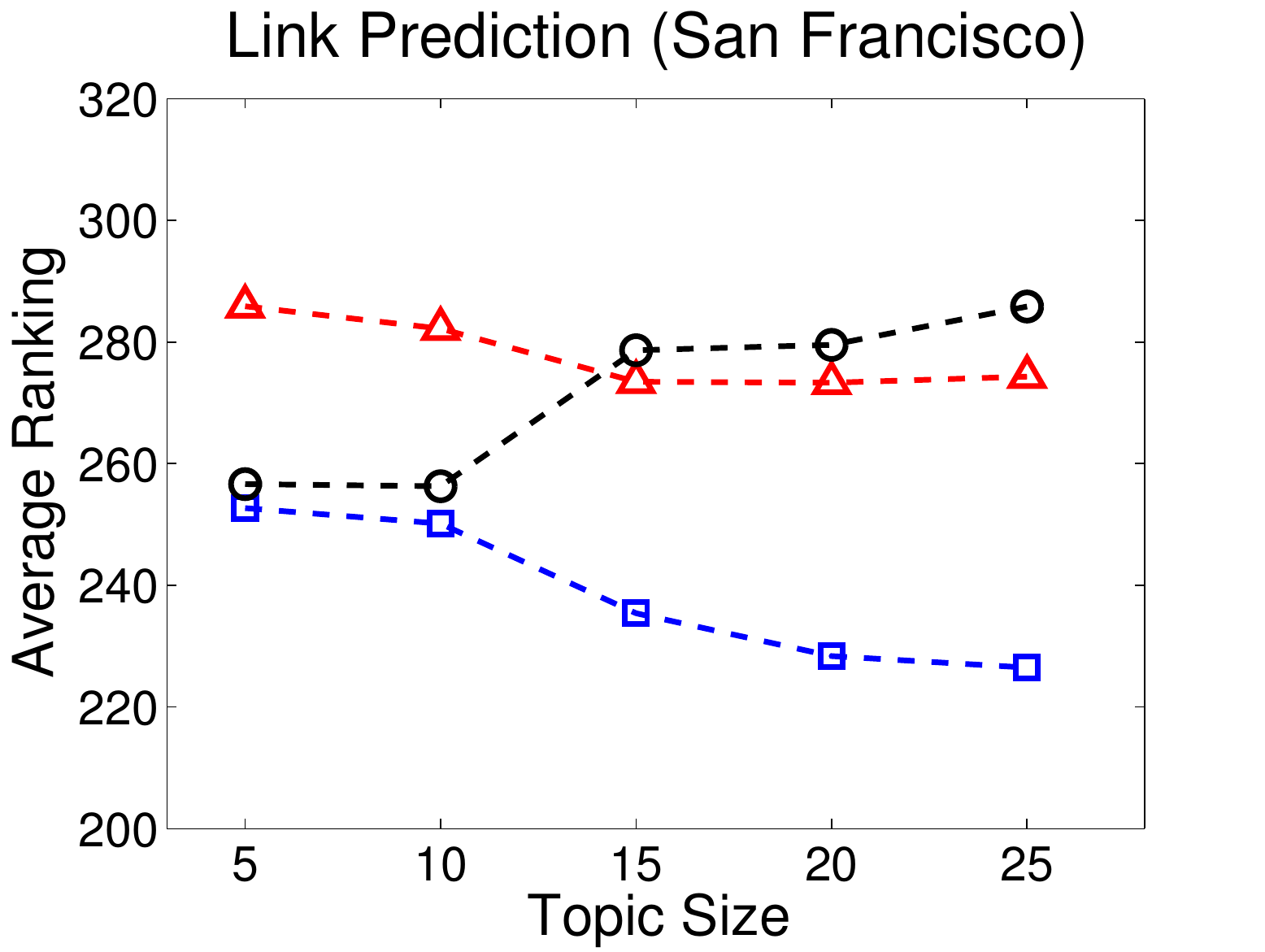} \includegraphics[width=0.234\textwidth]{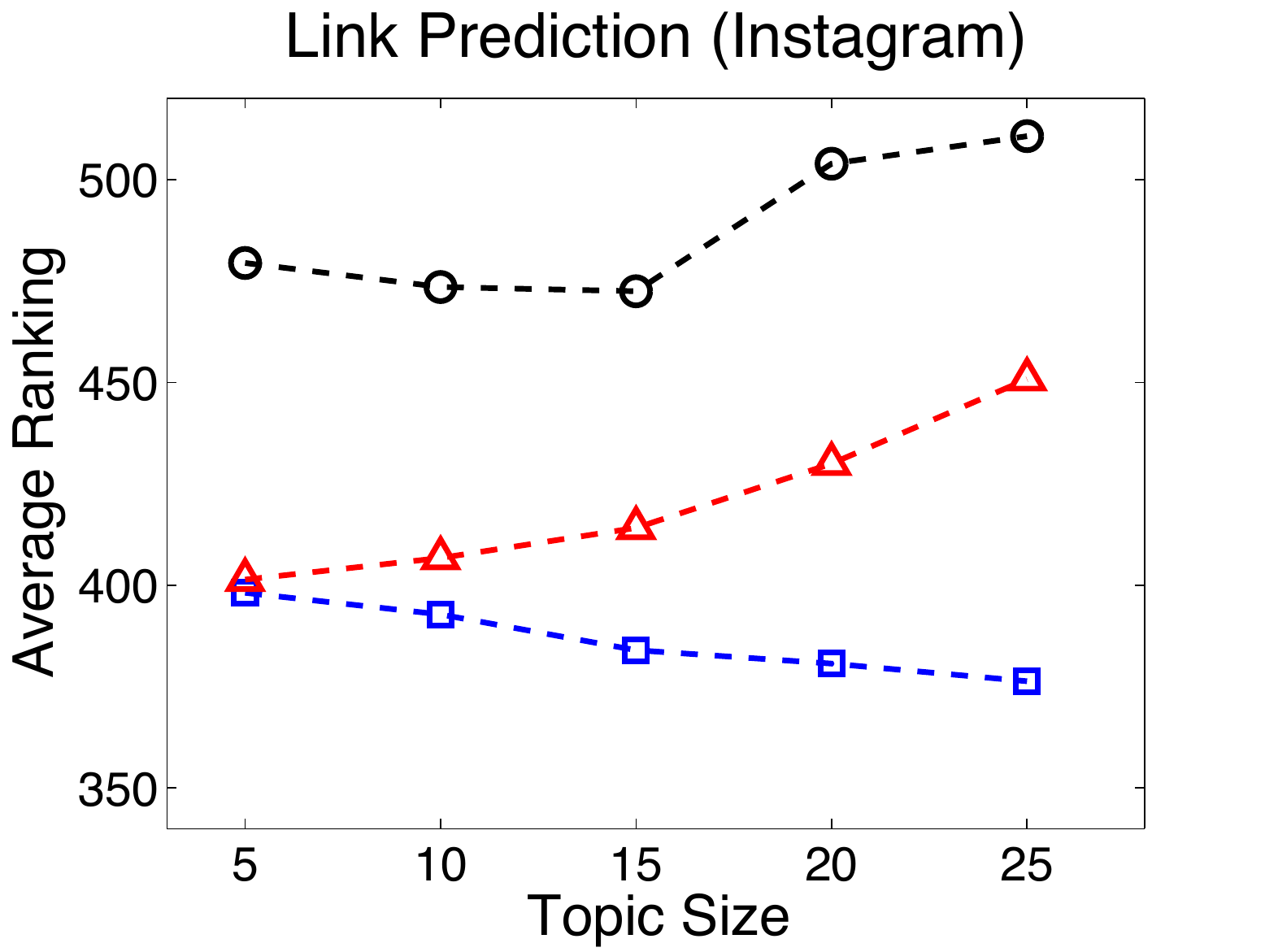} \includegraphics[width=0.234\textwidth]{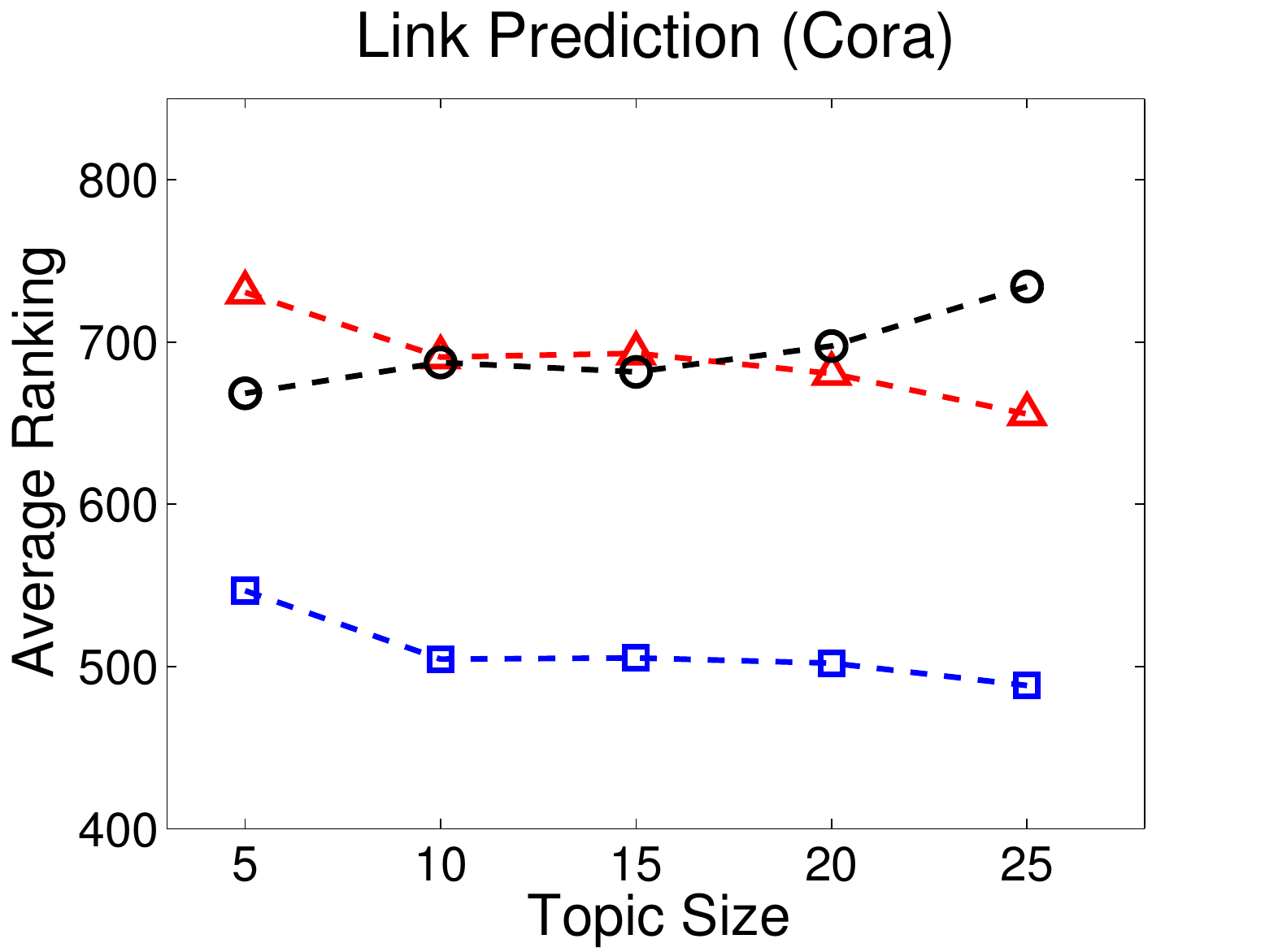}
  \includegraphics[width=0.234\textwidth]{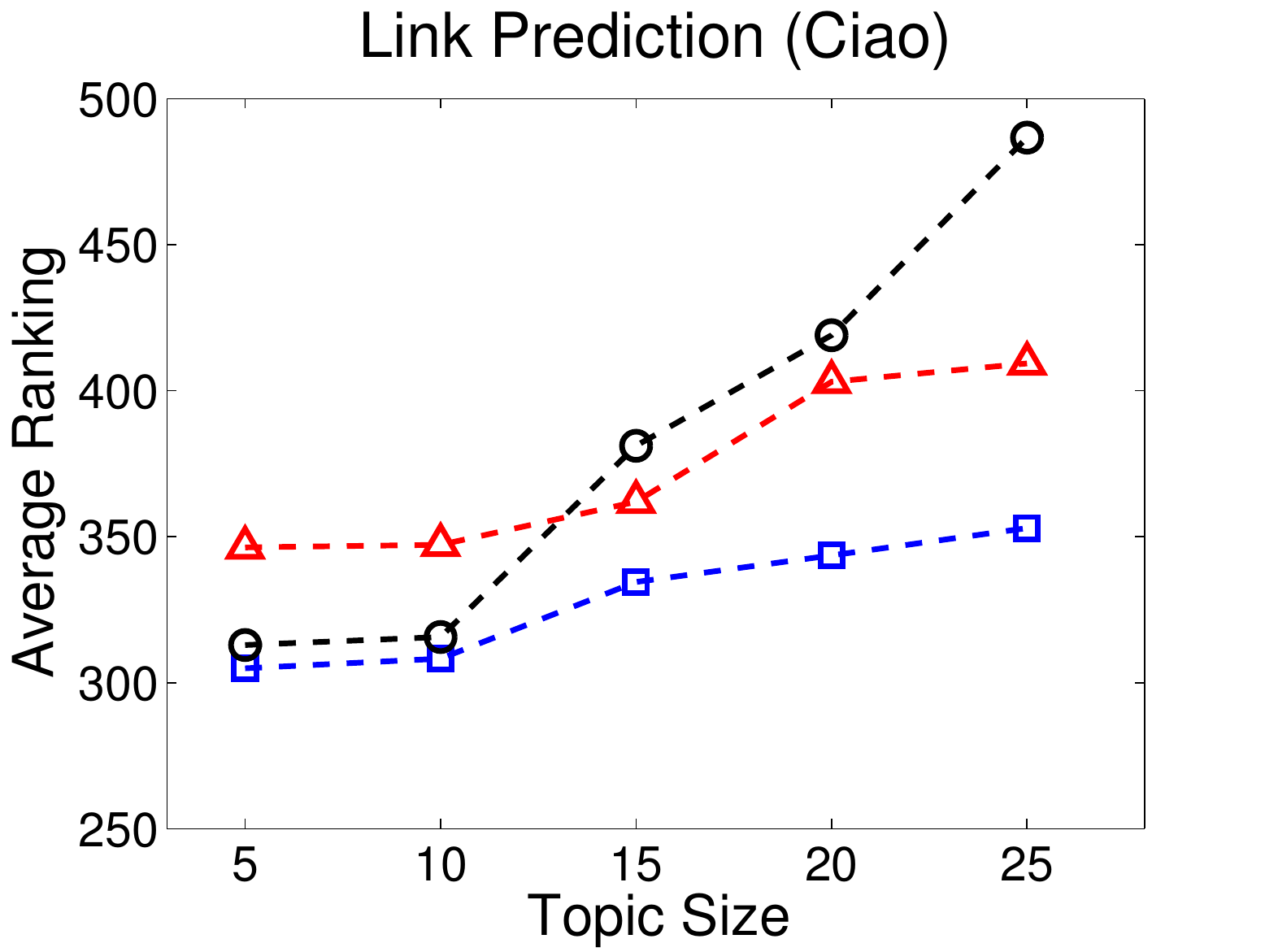}  \\
  \includegraphics[width=0.234\textwidth]{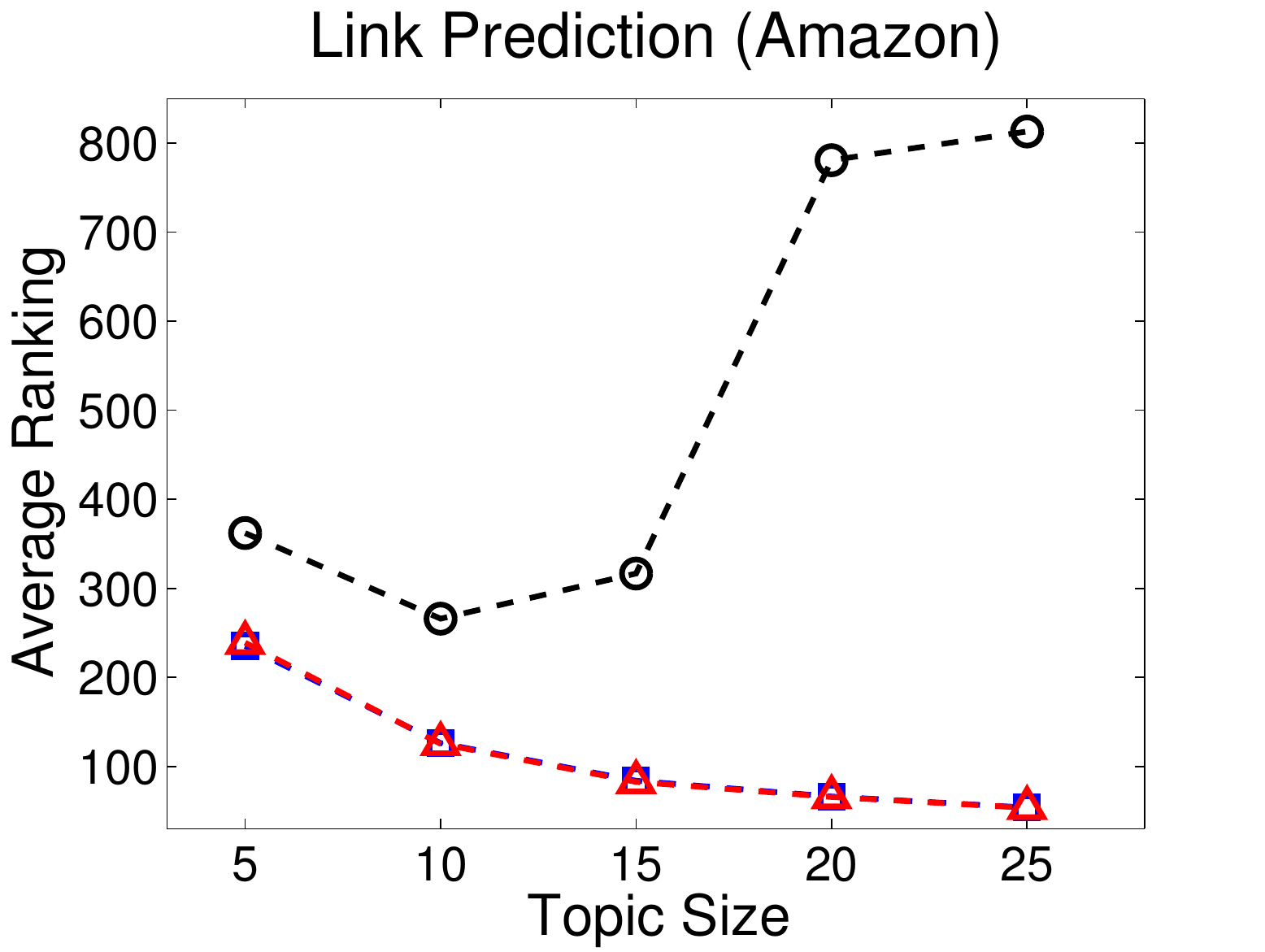} \includegraphics[width=0.234\textwidth]{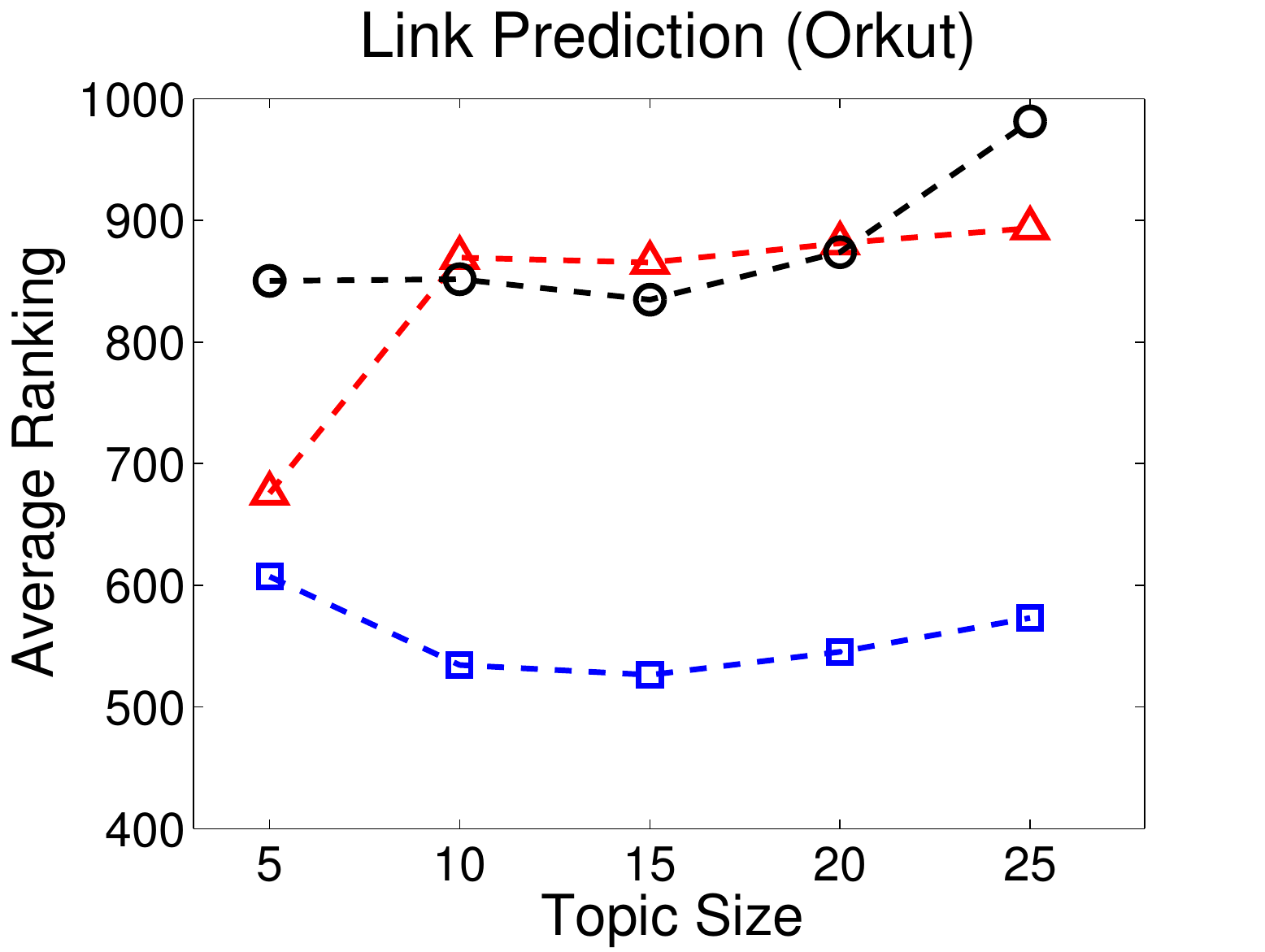} 
  
 \includegraphics[width=0.3\textwidth]{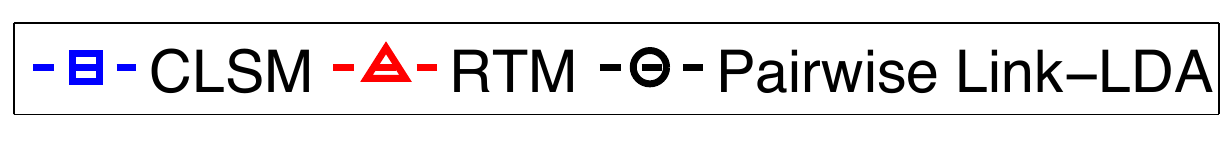}

\setlength{\abovecaptionskip}{-0.8pt}
\caption{Average ranking score of link prediction on six multi-modal datasets. Lower scores $\Leftrightarrow$ better performance.}
\label{fig:link}
\end{center}
\end{figure}

\begin{figure}[!t]
\vspace*{1mm}
\begin{center}

 \includegraphics[width=0.234\textwidth]{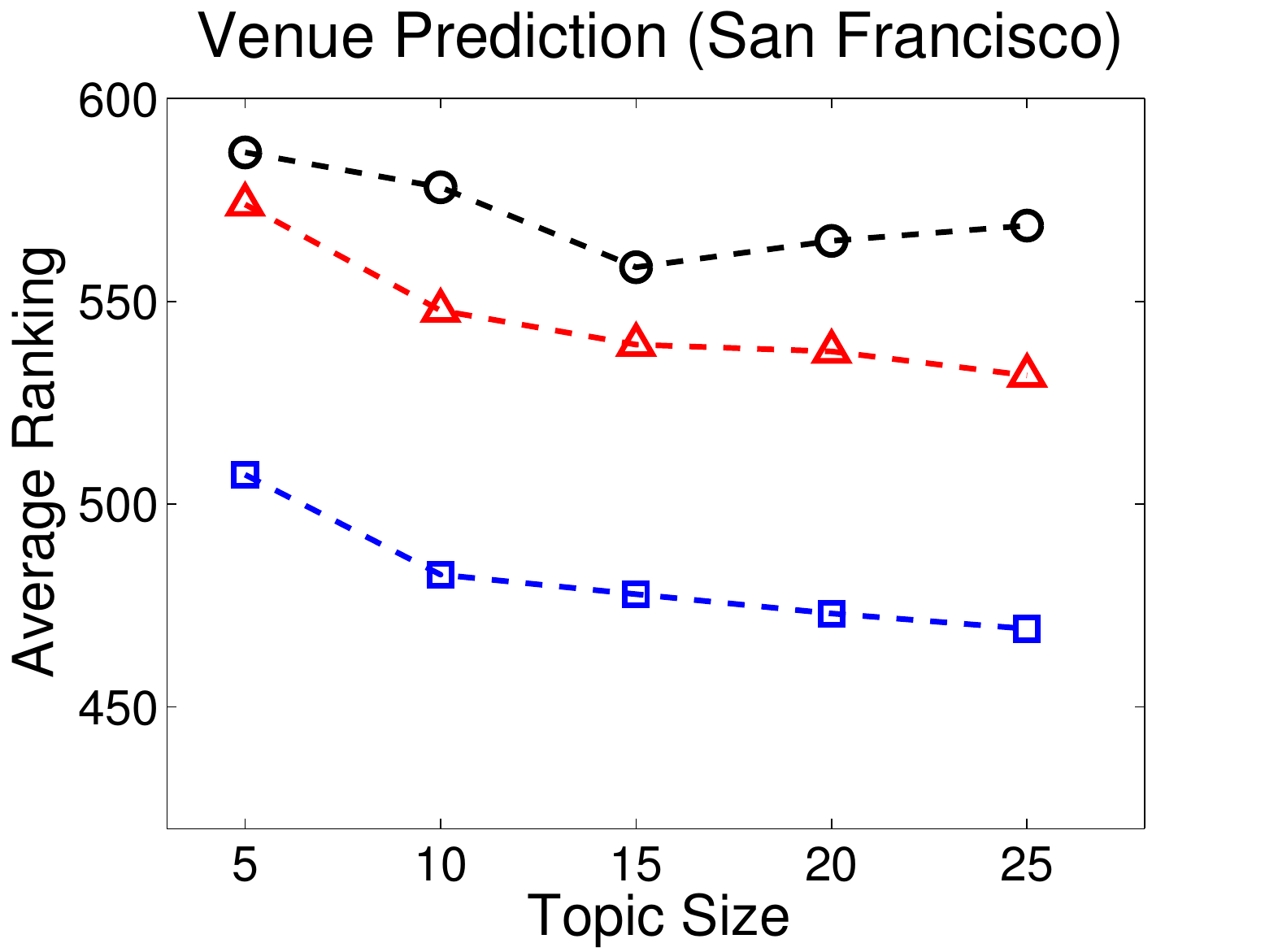}    \includegraphics[width=0.234\textwidth]{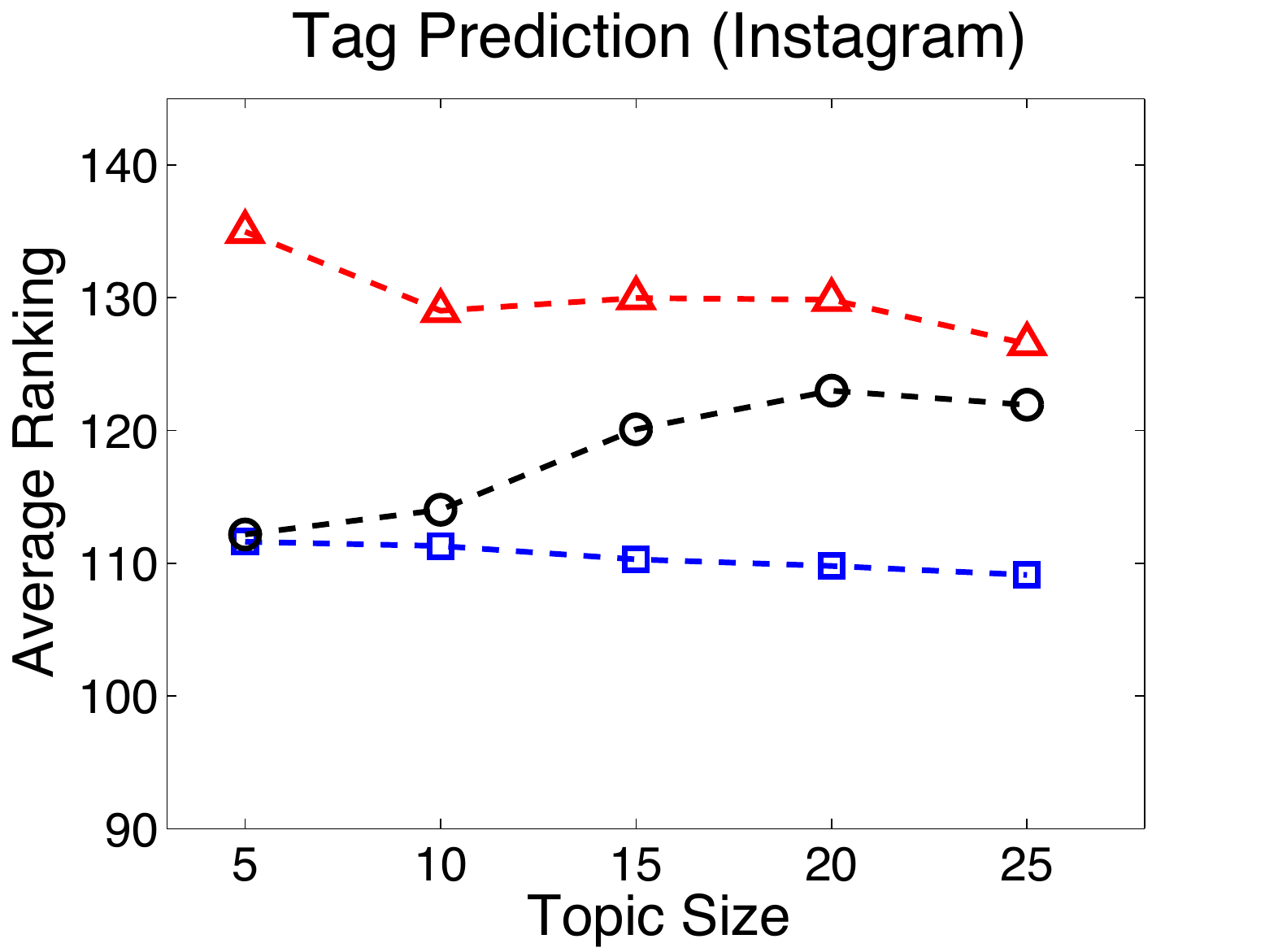}  \includegraphics[width=0.234\textwidth]{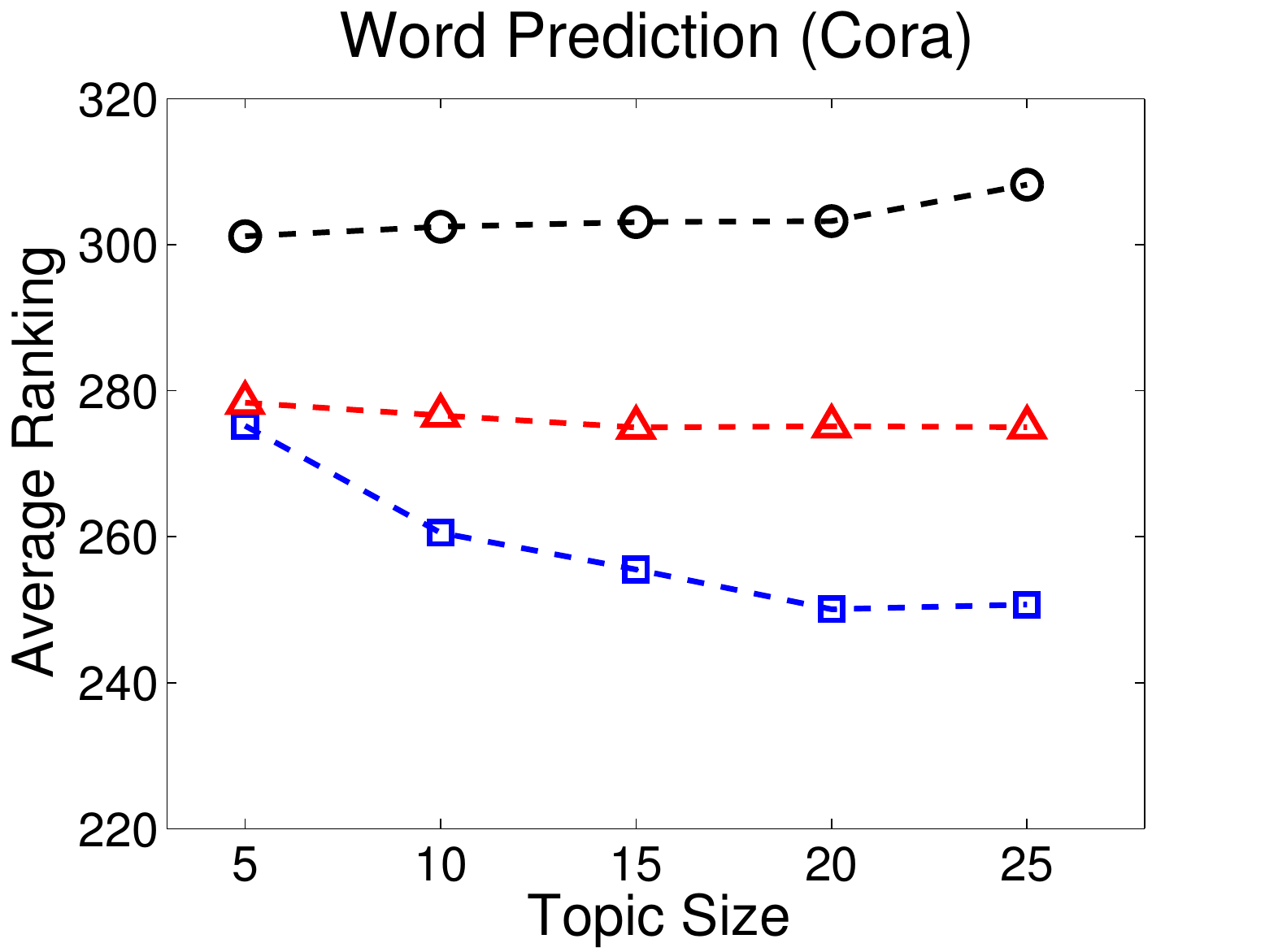}
   \includegraphics[width=0.234\textwidth]{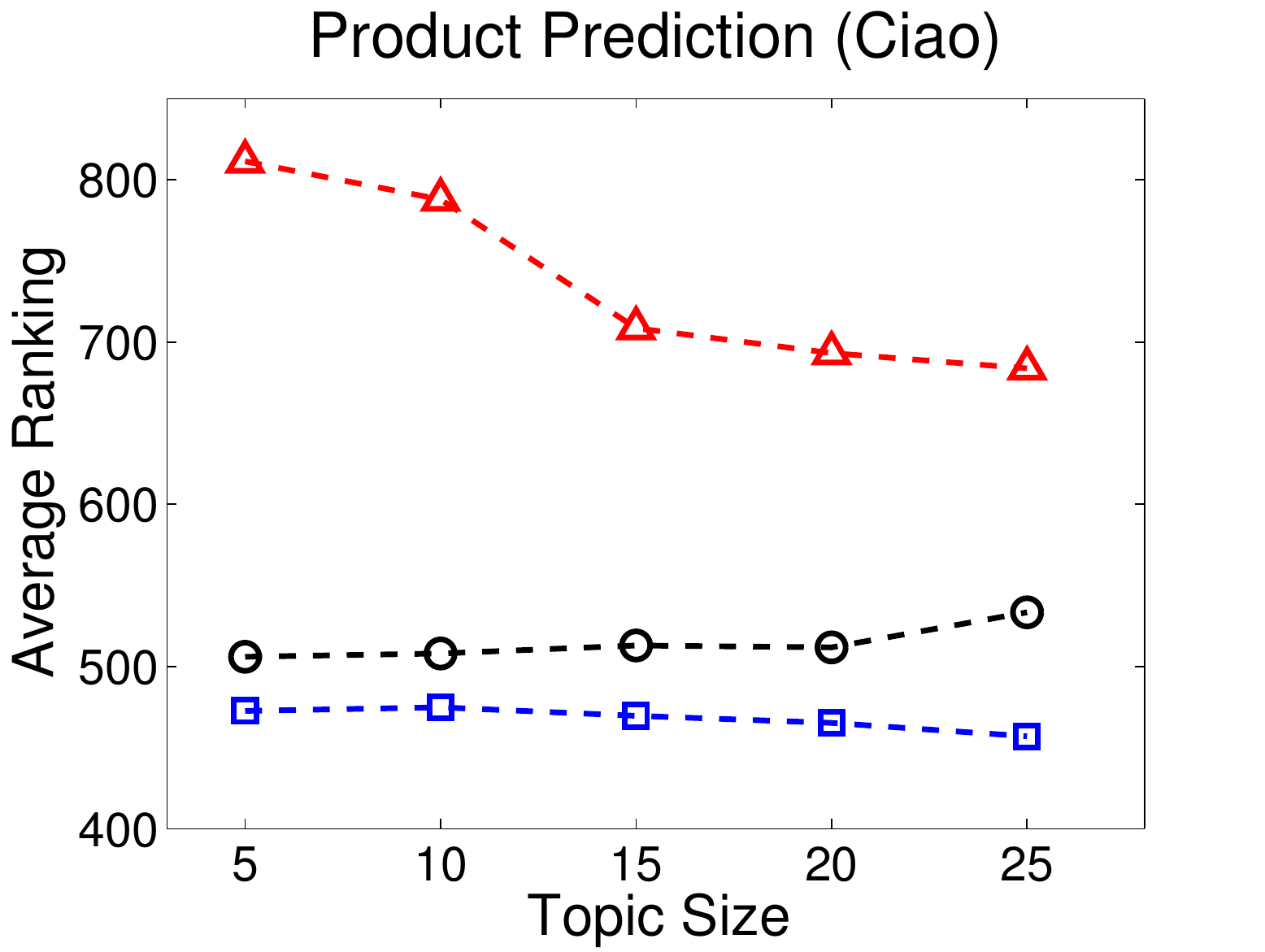} \\
  \includegraphics[width=0.234\textwidth]{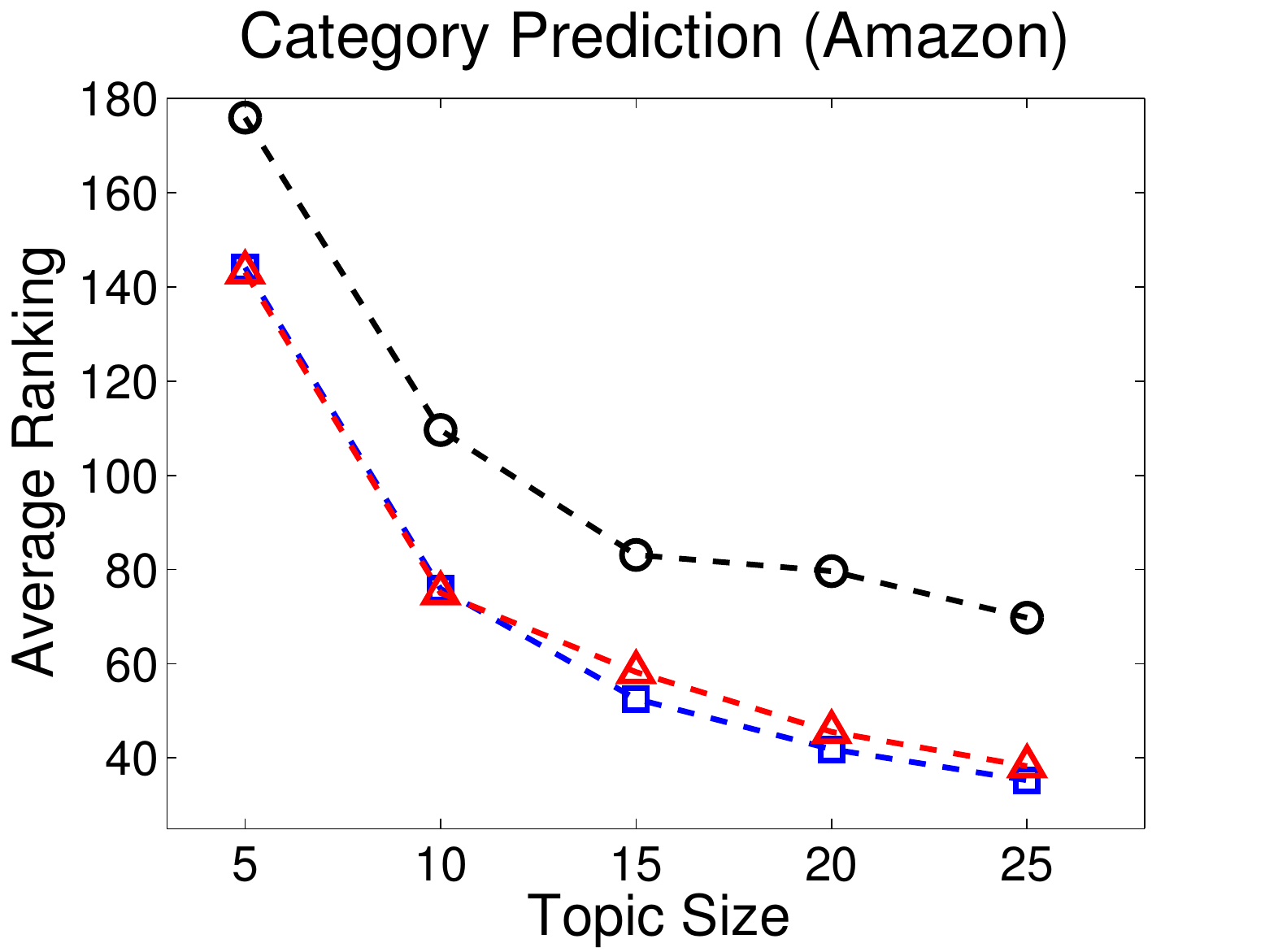}  \includegraphics[width=0.234\textwidth]{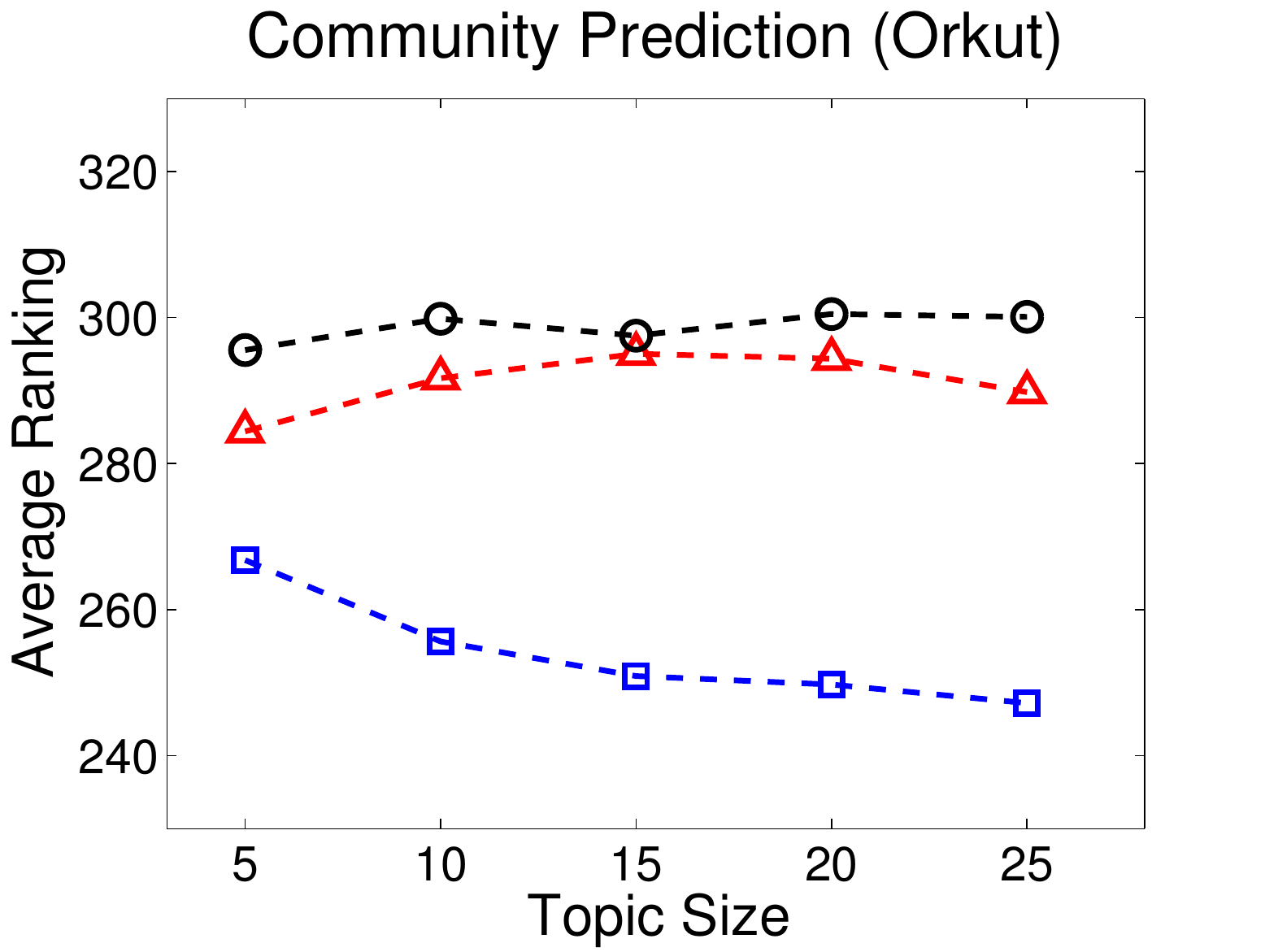} 
   
 \includegraphics[width=0.3\textwidth]{legend}
\setlength{\abovecaptionskip}{-0.8pt}
\caption{Average ranking score of attribute prediction on the six datasets.  Lower scores $\Leftrightarrow$ better performance.}
\label{fig:word}
\end{center}  
\end{figure}

\subsubsection{Task 2: Attribute Prediction} \label{sub:ap}
In this second task, we predict user attributes based on the attributes of other users while leveraging network information. For each node in the test set, we hide its attributes but assume that its social network links are known. Similarly to link recommendation, such predictions can be used to suggest products (or more general attributes) to a user. As in the link prediction problem, the average ranking of true attributes with $100\%$ recall is measured. The results are presented in Figure~\ref{fig:word}. Lower ranking scores imply that the actual attributes of a given node would appear ranked higher in the recommendation system. 

Our model provides better predictions on all datasets, outperforming the two baselines on the attribute prediction task as well. The prediction accuracy scores (measured by AUC) for the three models are reported in Table~\ref{tb:AUC}: CLSM yields a performance improvement on average of 5.14\% and 10.16\% over Pairwise Link-LDA and RTM respectively. Our model is also the only one that exhibits a constant improvement as the number of topics grows: the average ranking scores produced by CLSM monotonically decreases in all datasets.

All models do well with \emph{Amazon} data: a plausible hypothesis is that \emph{Amazon} uses co-purchasing patterns to perform recommendations, and therefore the high AUC scores capture the ability of all models to reproduce such feature.
We also note that Pairwise Link-LDA performs better than RTM on \emph{Gowalla} venues prediction, and on \emph{Amazon} category prediction.
This may be due to the flexibility that aMMSB allows in the presence of social network interactions where people tend to be assigned to multiple communities. 

The prediction tasks we presented also unveiled some limits and challenges of these models: for example, the improvement aMMSB yields tend to reduce in {\it Cora}, possibly as a consequence of the presence of few topics in the documents. Finally, \emph{Gowalla} data suggest that location-based social networks pose some serious challenges, yielding an accuracy of only 70\% and 75\% for link and attribute prediction respectively. We will explore in further details these two scenarios in the following in-depth analysis.

\begin{table}[t]
\small
\caption{Accuracy (AUC) in link and attribute prediction.}
\label{tb:AUC}
\begin{center}
\begin{tabular}{| c | c  ||  c | c |}
\hline
\multicolumn{4}{ |c| }{AUC scores with $K = 15$} \\
\hline  
Dataset & Model & Link AUC &Attribute AUC\\  \hline
\multirow{3}{*}{\emph{Gowalla} (SF)} & PL-LDA  & 0.6278  & 0.7154 \\
 & RTM & 0.6300 & 0.6746\\
 & \bf{CLSM} & \bf{0.7010} & \bf{0.7429} \\  \hline
\multirow{3}{*}{{\it Cora}} & PL-LDA & 0.6341 &0.7928 \\
 & RTM & 0.6509 &0.7971\\
 & \bf{CLSM} & \bf{0.7570} & \bf{0.8203}\\ \hline
\multirow{3}{*}{{\it Ciao}} & PL-LDA & 0.7038 &0.7636 \\
 & RTM & 0.6738 &0.6850\\
 & \bf{CLSM} & \bf{0.7532} & \bf{0.7929}\\ \hline
\multirow{3}{*}{{\it Amazon}} & PL-LDA & 0.8565 &0.9661 \\
 & RTM & 0.9586 &0.9752\\
 & \bf{CLSM} & \bf{0.9686} & \bf{0.9765}\\ \hline
 \multirow{3}{*}{{\it Orkut}} & PL-LDA & 0.8837 &0.6884 \\
 & RTM & 0.8836 &0.6580\\
 & \bf{CLSM} & \bf{0.9031} & \bf{0.7890}\\ \hline
\multirow{3}{*}{\emph{Instagram}} & PL-LDA & 0.6385 &0.7113 \\
 & RTM & 0.6971 &0.6598  \\
 & \bf{CLSM} & \bf{0.7791 } & \bf{0.7401 }\\ \hline
\end{tabular}
\end{center}
\end{table}

\subsection{Case Studies} \label{sub:scenarios}
\textbf{Case Study 1: Analysis of Cora.} In this data set, each document is assigned to one of the following categories: \emph{Neural Networks}, \emph{Rule Learning}, \emph{Reinforcement Learning}, \emph{Probabilistic Methods}, \emph{Theory}, \emph{Genetic Algorithms}, and \emph{Case Based in Machine Learning}. Figure~\ref{fig:totalcora} shows the count of documents for each category: over half of the documents belong to either \emph{Neural Networks}, \emph{Probabilistic Methods}, or \emph{Genetic Algorithms}. Without access to this category information, our model only uses the contained words and link information to infer the topics of the documents. We here investigate how these categories match the topics we inferred. For each document, we specify a major topic and collect all documents in the corpus related to that topic. Figure~\ref{fig:pie_subplot} shows the percentage of categories for each of the 10 topics, where we have used $K=10$ (number of topics) for illustrative purposes. The documents which have topic 1 as major topic are dominated by \emph{Neural Networks}. We can also see how many topics favor specific categories over others. Some of the topics contain mixtures of labels: this is due to the loose separation between the category labels. When the categories are closely related, this tends to be captured in the pie chart of our inferred topics forming a mixture. With higher number of total topics (e.g., $K=25$), we can observe a better separation of labels over topics. However, even with small size of $K$, we can still obtain meaningful insights such as how different categories relate each other (i.e., whether two categories are often captured together in a topic). 

\begin{figure} 
        \centering
        \begin{subfigure}[b]{0.4\textwidth}
                \includegraphics[width=\textwidth,clip=true,trim=10 35 0 0]{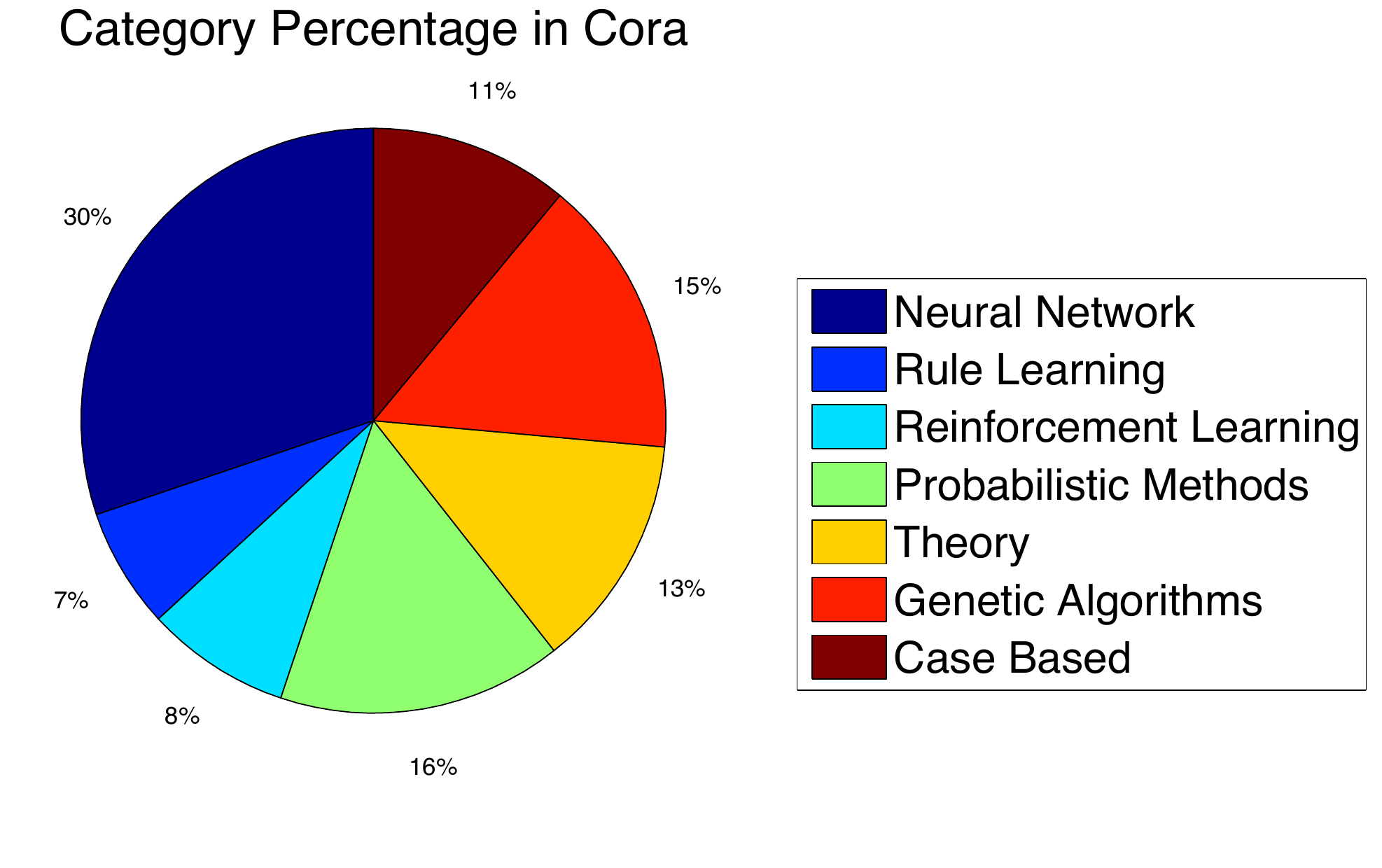}
                \caption{}
                \label{fig:totalcora}
        \end{subfigure}
        \begin{subfigure}[b]{0.48\textwidth}
                \includegraphics[width=\textwidth]{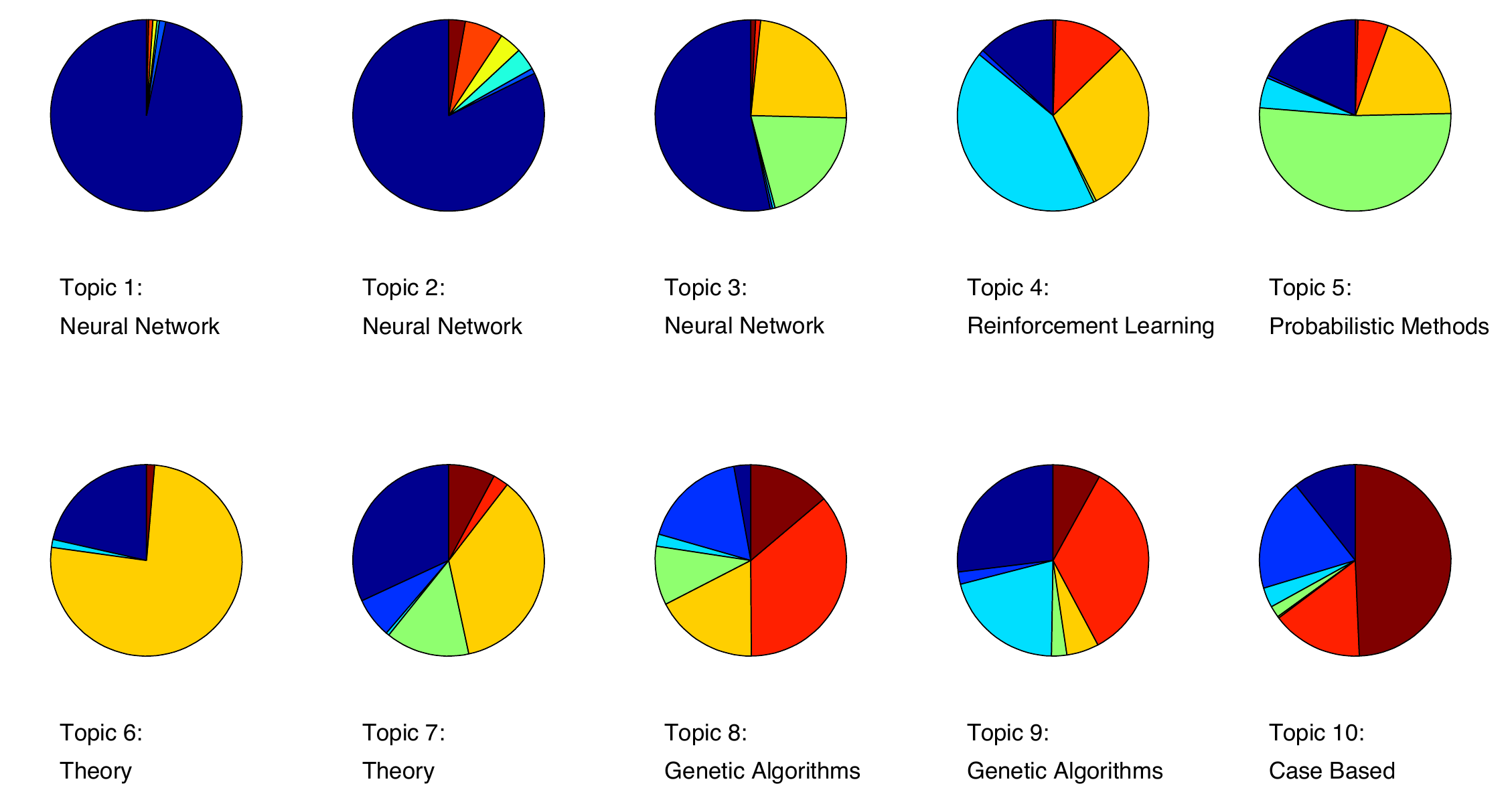}
                \caption{}
                \label{fig:pie_subplot}
        \end{subfigure}
        \caption{(a) The percentage of known categories in \emph{Cora}. (b) The percentage of the categories for the collection of documents of each topics ($K=10$). }
        \label{fig:pie} 
\end{figure}

\begin{figure}[!t]
        \centering
        \begin{subfigure}[b]{0.35\textwidth}
                \includegraphics[width=\textwidth,clip=true,trim=5 0 5 5]{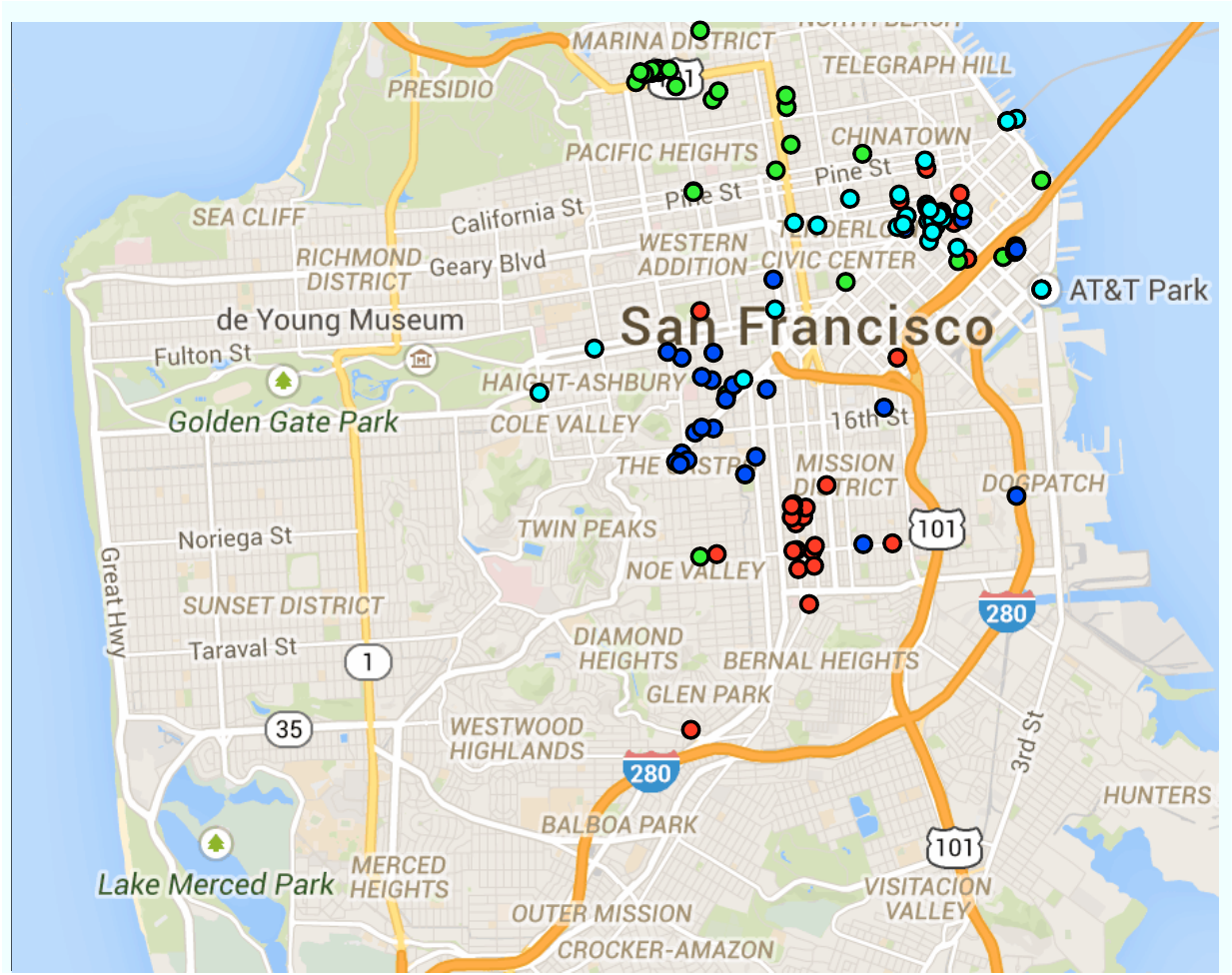}
                \label{fig:SF}
        \end{subfigure}
              \caption{The top 30 attended venues from the four most popular clusters are displayed in different colors. The emergence of neighborhoods appear evident. 
              }
        \label{fig:SF_map} 
\end{figure}

\begin{figure*}[!t]
\vspace*{1mm}
\begin{center}
\includegraphics[width=0.028\textwidth]{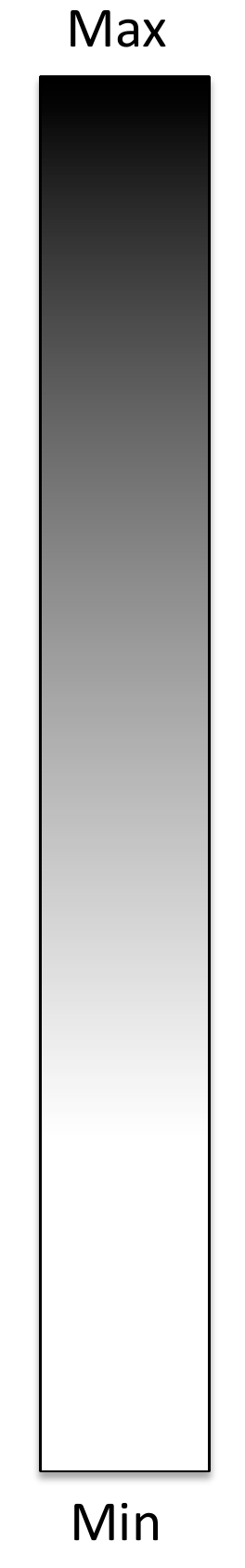}  \includegraphics[width=0.235\textwidth]{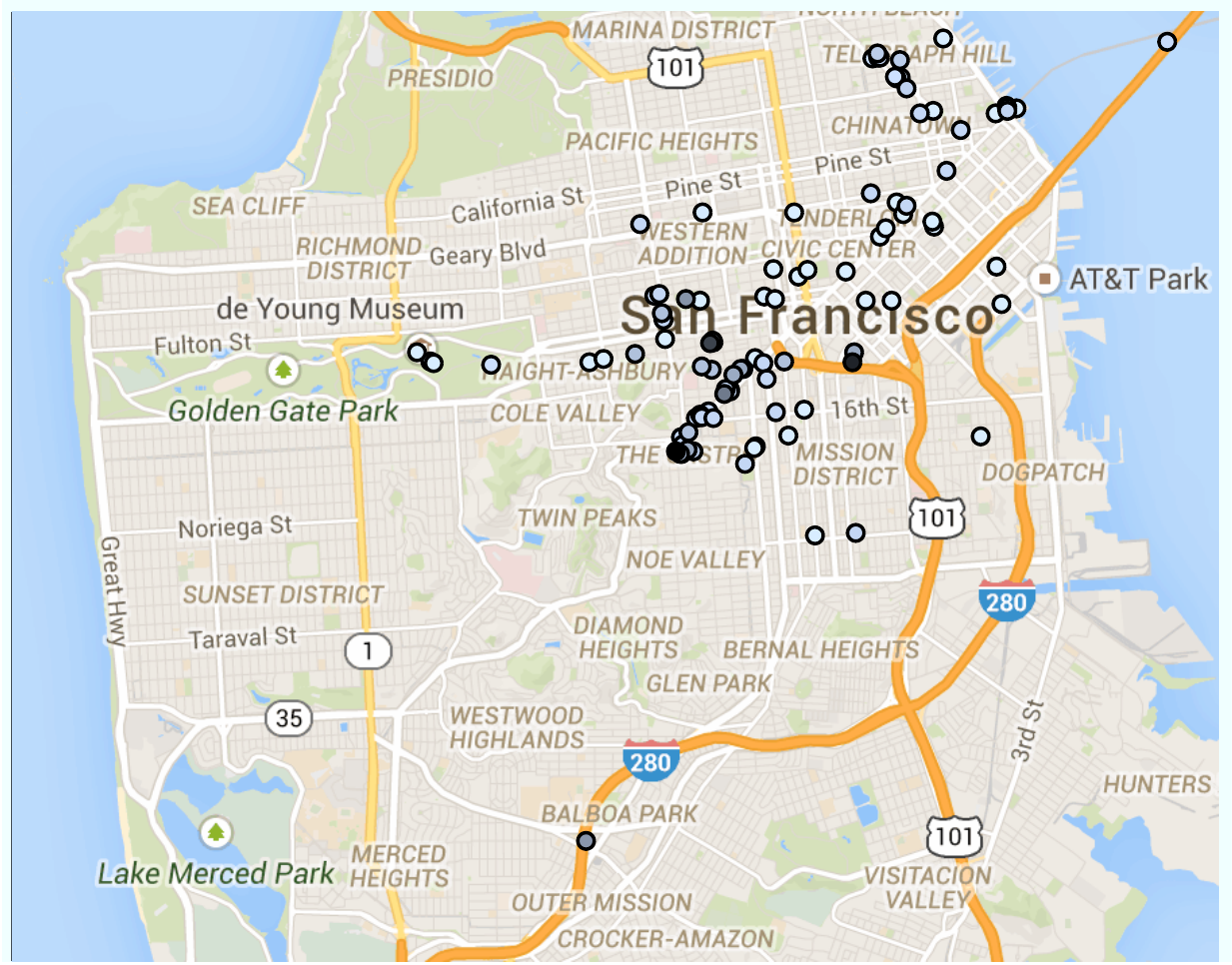} \includegraphics[width=0.235\textwidth]{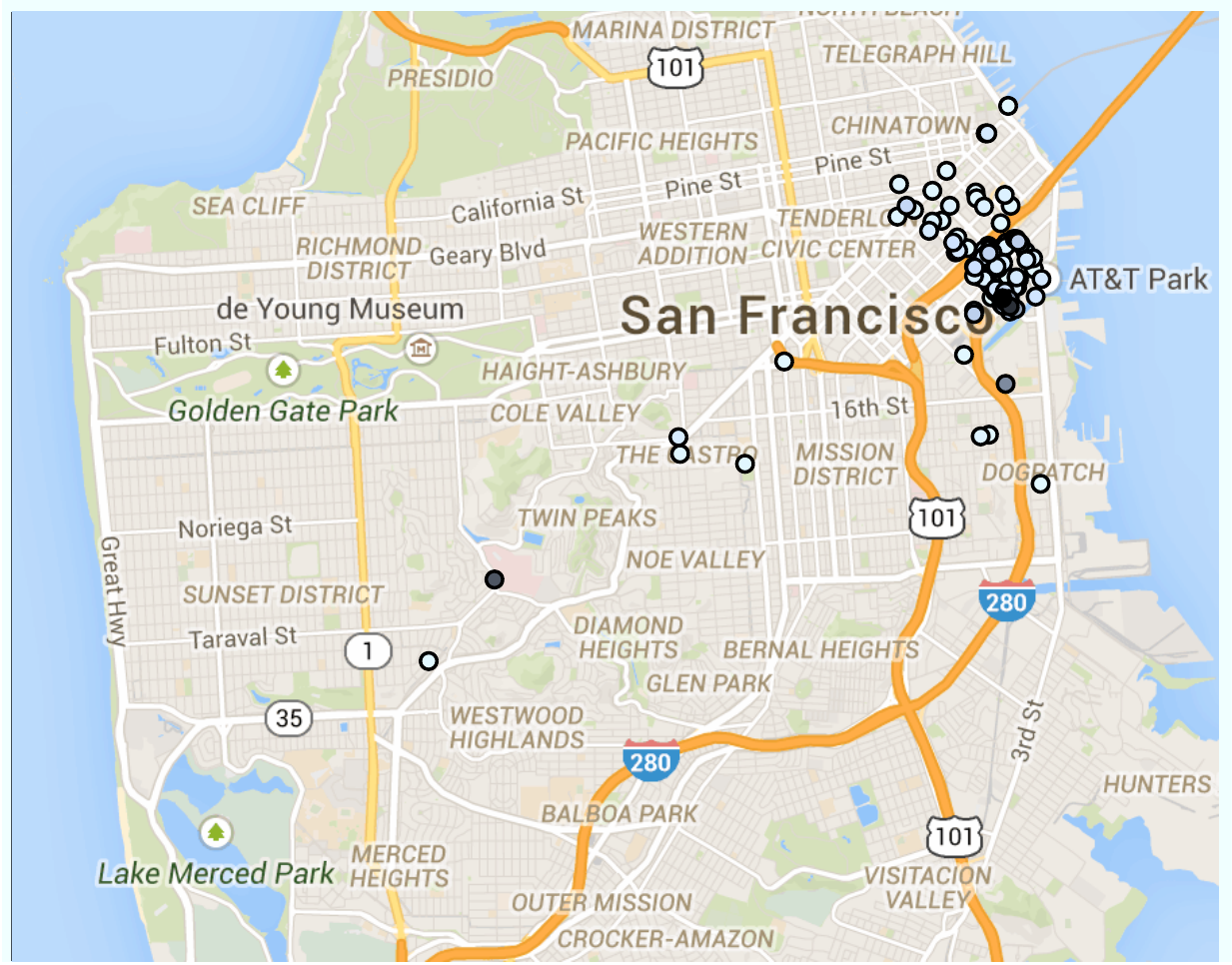}  \includegraphics[width=0.235\textwidth]{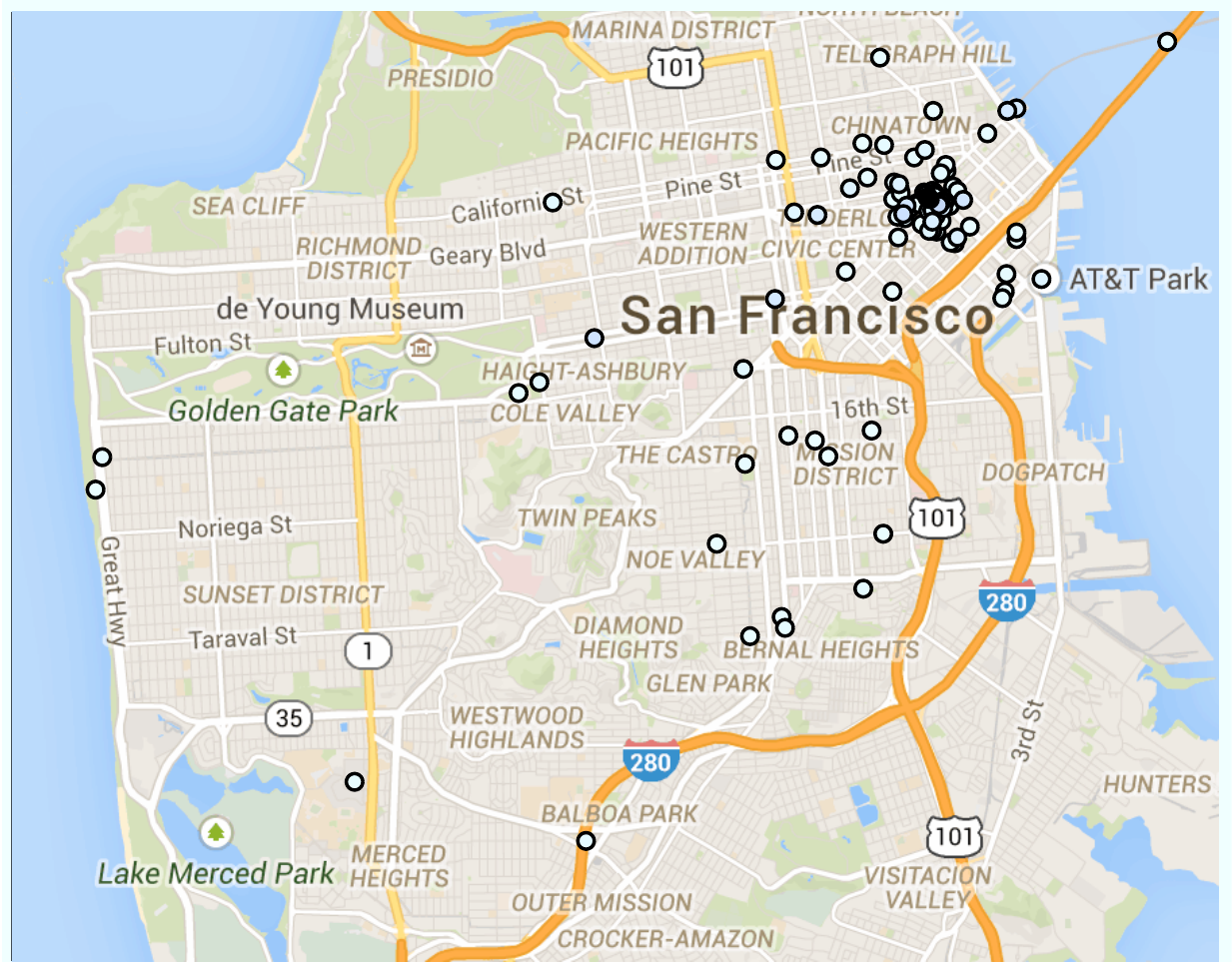} \includegraphics[width=0.235\textwidth]{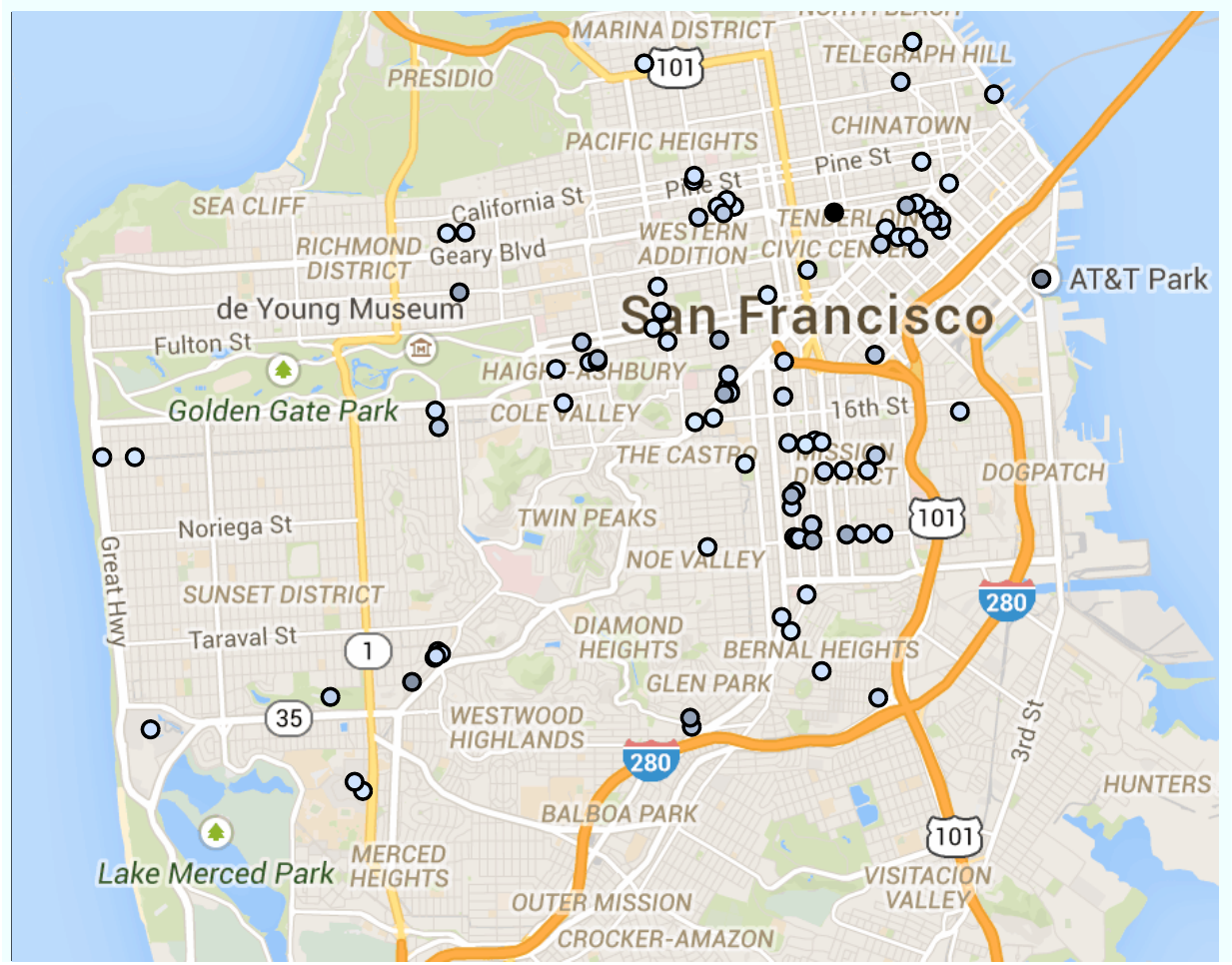}
 \caption{Maps of San Francisco's four most attended clusters, and their top 100 attended venues. The three leftmost maps show clusters that mostly capture physical proximity. However, the rightmost map shows that CLSM can also capture other dimensions: in such a case for example we can see many tourist spots emerging due to the social nature of human traveling.}
 \label{fig:location}
 \end{center}
 
\end{figure*}

\smallskip\textbf{Case Study 2: Analysis of Gowalla.} 
To the best of our knowledge, our work is one of the first attempts at finding the joint topic space in a location-based social network. Previous studies modeled social networks and user mobility patterns~\cite{FM,geo_coincidence,FF,Wang:2011:HMS:2020408.2020581}. For example, Wang \emph{et al.}~\cite{Wang:2011:HMS:2020408.2020581} measured the co-visitation frequency to infer social interactions. However, their model suffers from capturing ties between nodes which often exhibit no co-occurrence in locations within physical proximity. 
We previously introduced a latent space representation of venues to predict social ties~\cite{4691} using \emph{agglomerative information bottlenecks} to cluster venues, and compared it to other models including one based on LDA. However, that approach only clusters venues, while ideally one would prefer to cluster venues and users simultaneously.
Here, CLSM applied to \emph{Gowalla} data finds the joint latent space that simultaneously describes the friendship between users and their  {\it check-in} behavior (i.e., venues they attend). Using the geo-coordinates provided for each venue, we further investigate the geographical distribution of  popular venues. Even though we have not explicitly included the geo-coordinates in our inference algorithm, we may reasonably assume that geo-clusters (neighborhoods) should be reflected in our inferred topics (clusters), since some people make friends with others who live or work close-by. 
Our analysis focuses on San Francisco: in Figure~\ref{fig:SF_map} we produce the results of the attribute prediction task with CLSM by setting the total number of topics (clusters) to $K=25$.
After selecting the four most popular clusters, we visualize the overall top 30 most attended venues: we can appreciate how the four clusters appear geographically well separated. Let us analyze them more into detail: Figure~\ref{fig:location} this time shows the top 100 popular venues for each of the four clusters above. From left to right, the first three maps show that indeed the visited venues are affected by physical distances, and we can observe the emergence of neighborhoods. However, the rightmost map shows that our model is also able to capture clusters not necessarily representing physical closeness: further inspection reveals that this cluster represents tourist spots that are often attended by San Francisco's visitors. None of the other models was able to capture simultaneously such diverse patterns: CSLM overcomes this limit by leveraging the multi-modal nature of social data.

\section{Conclusion}   \label{Conclude}
Techno-social systems exhibit a remarkable amount of complexity, capturing not only the  interactions among users but also their attributes and behaviors along multiple dimensions. For example, in location-based social networks users can be concurrently described by their mobility patterns, their activities, their preferences, and of course by their social links. In online social networks users produce and consume contents, link each other, join groups, etc. 
Recent studies illustrated that these modalities taken independently cannot capture the multiple facets of user activity and behavior.
As a result, our ability to effectively model, design, analyze and improve such systems substantially depends on the possibility of leveraging the abundance of rich contextual information. 
In summary, very many practical problems commonly occurring when designing or analyzing socio-technical systems would greatly benefit from a multi-modal modeling framework. 

To address these challenges, in this paper we proposed the Constrained Latent Space Model (CLSM), which employs a multi-modal paradigm to simultaneously describe social network information and user behavioral data using a latent space representation.
To describe the network generative process, CLSM leverages Mixed Membership Stochastic Blockmodels (MMSB) that captures mixed memberships, nodes that may belong to more than one community at the same time. The latent space is inferred via Latent Dirichlet Allocation (LDA).
One remarkable characteristics of CLSM is that it introduces a constrain that enforces MMSB and LDA to overlap on the same latent space without loss of flexibility. To tame the algorithmic complexity of such a task, we suggested an efficient inference strategy based on Variational Expectation Maximization, which scales linearly with the size of the network. Experiments with synthetic data illustrate the advantage of such an approach.

To further show the flexibility of our framework, we designed two evaluation tasks inspired by prediction problems commonly occurring with real techno-social systems: \emph{(i)} a link prediction experiment that aims at reconstructing the missing links among users given available attributes or behavior data; \emph{(ii)} an attribute prediction experiment where we infer user attributes (or behaviors) leveraging social interaction data.
A rigorous evaluation of our model against two existing approaches (Pairwise Link-LDA and Relational Topic Model) illustrates the superior performance of CLSM. 
The benchmark performed on six different multi-modal social datasets includes location-based social networks, social sharing platforms, etc.
We reported the increments in prediction accuracy (measured by AUC) yielded by CLSM with respect to the state of the art on all the evaluation scenarios: CLSM outperforms Pairwise Link-LDA on average by 12.56\% and 5.14\% in link and attribute prediction, respectively, and it surpasses RTM by 9.06\% and 10.16\% on the same tasks. We further discuss two scenarios in details, describing the insights obtained adopting our multi-modal framework that would have not been otherwise possible.

The main appealing features of the proposed approach include its excellent scalability and its ability to handle various types of multi-modal datasets with relational information. It is worth noting that, although we here report only analyses of bimodal networks, CLSM can be adopted with data capturing any number of modalities. For such a reason, in the future we plan to use our model on datasets with even more modalities (e.g., multi-layered attributes, or combinations of different types of relational information between nodes).

\bibliographystyle{abbrv}
\balance 
\bibliography{main}

\end{document}